\documentclass[preprint,11pt]{elsarticle}
\usepackage[english]{babel}
\usepackage[latin1]{inputenc}
\usepackage[T1]{fontenc}
\usepackage{graphicx,color}
\usepackage{amsmath}
\usepackage{amssymb}
\usepackage{array}
\usepackage[margin=1.8cm]{geometry}
\begin{document}

\begin{frontmatter}

\title{Detailed analysis of the effects of stencil spatial variations with arbitrary high-order finite-difference Maxwell solver. }

\author[label1,label2]{H. Vincenti\corref{cor1}}
\ead{hvincenti@lbl.gov}

\author[label1]{J-L Vay}
\ead{jlvay@lbl.gov}

\address[label1]{Lawrence Berkeley National Laboratory, 1 cyclotron road, Berkeley, California, USA}

\address[label2]{Lasers Interactions and Dynamics Laboratory (LIDyL), Commissariat à l'Energie Atomique, Gif-Sur-Yvette, France}

\cortext[cor1]{Corresponding author}

\begin{abstract} 

Due to discretization effects and truncation to finite domains, many electromagnetic simulations present non-physical modifications of Maxwell's equations in space that may generate spurious signals affecting the overall accuracy of the result. Such modifications for instance occur when Perfectly Matched Layers (PMLs) are used at simulation domain boundaries to simulate open media. Another example is the use of arbitrary order Maxwell solver with domain decomposition technique that may under some condition involve stencil truncations at subdomain boundaries, resulting in small spurious errors that do eventually build up. In each case, a careful evaluation of the characteristics and magnitude of the errors resulting from these approximations, and their impact at any frequency and angle, requires detailed analytical and numerical studies. To this end, we present a general analytical approach that enables the evaluation of numerical discretization errors of fully three-dimensional arbitrary order finite-difference Maxwell solver, with arbitrary modification of the local stencil in the simulation domain. The analytical model is validated against simulations of domain decomposition technique and PMLs, when these are used with very high-order Maxwell solver, as well as in the infinite order limit of pseudo-spectral solvers. Results confirm that the new analytical approach enables exact predictions in each case. It also confirms that the domain decomposition technique can be used with very high-order Maxwell solver and a reasonably low number of guard cells with negligible effects on the whole accuracy of the simulation. 
\end{abstract}

\end{frontmatter}

\newpage

\section{Introduction}


Very high-order Finite-Difference Time-Domain (FDTD) Maxwell solver and, in the infinite order limit pseudo-spectral solvers, are methods of choice for solving three-dimensional electromagnetic problems that require very high accuracy over a large band of frequencies. In Particle-In-Cell (PIC) simulations of laser-plasma mirror interactions for instance, a good dispersion relation is needed over a large band of frequencies and angles for the modeling of relativistic harmonic generation \cite{Thaury2007,Vincenti2014}. In PIC simulations of laser-plasma acceleration, numerical dispersion induces numerical Cherenkov effects \cite{Godfrey1974504,Greenwood2004665} that can be highly detrimental for the modeling of ultra relativistic beams and plasmas. In general, the mitigation of all these effects requires the use of very high spatial and temporal resolutions which, with current low order FDTD solvers, sometimes prevent parametric studies of large parameter space at scale in full three dimensions. In the case of numerical Cherenkov effects, it was shown analytically and numerically that pseudo-spectral solvers are generally more stable than standard second-order schemes \cite{Godfrey2014}. The use of very high-order or pseudo-spectral solvers can significantly decrease the resolution needs and increase the overall stability for a given accuracy and can thus enable realistic 3D PIC simulation studies that are otherwise not practical. 

Despite significant advantages in terms of accuracy and memory-minimization \cite{Boyd1989}, high-order and pseudo-spectral solvers have however not been widely adopted so far for large-scale simulations on massively parallel supercomputers because of their low parallel scalability, which is a direct consequence of their spatial non-locality (large stencils). Indeed, these solvers commonly use Fast Fourier Transform (FFT)-based algorithms that require global inter-processor communications in the computation of Fourier transforms, limiting their scaling to a few thousands of cores.  This is preventing the use of pseudo-spectral solvers in very computationally demanding 3D plasma electromagnetic simulations that usually mandate several hundred of thousands of cores. 

Recently, a new method \cite{Vay2013} for solving time-dependent problems (e.g. Maxwell's wave equations) proposed to apply the cartesian domain decomposition technique currently used with low order FDTD solvers to very high-order pseudo-spectral solvers. As in the case of low-order schemes, the simulation domain is divided into several subdomains and Maxwell's equations solved locally on each subdomain using local convolution (e.g. FDTD), or local FFT's when the order $p$ becomes very large. The fundamental argument legitimating this method is that physical information cannot travel faster than the speed of light. Choosing large enough guard regions should therefore ensure that spurious signal stemming from stencil truncations at subdomains boundaries is practically limited to the guard regions and that only a negligible fraction reenters the simulation domain after one time step. One potential drawback of this approach is the non-locality of Gibbs oscillations arising when a signal is truncated at the edges of the guard regions. Indeed, depending on the size of guard regions, errors stemming from stencil truncations may affect the entire simulation domain and instabilities could build up. Studying the impact of stencil modifications/truncations on the overall accuracy of the simulation is thus crucial to validate the use of arbitrary order schemes with domain decomposition techniques. 




%
\subsection{Goals and outline of this study}

The goal of this study is to provide a very general analytical approach that enables the prediction of the total amount of error coming from stencil modifications in 3D electromagnetic simulations. For a fixed simulation configuration, this approach can be used to compute the optimal choice of numerical parameters (e.g. solver order $p$, space and time steps, number of guard cells) that will compute the solution in a minimum time and with a guaranteed accuracy. Our model will be especially relevant to predict errors when cartesian domain decomposition is used along with very high order/pseudo-spectral solvers in electromagnetic simulations. It is also applicable to high-accuracy prediction of the performance of PMLs with high-order or pseudo-spectral solvers.

The study is divided in three sections: 

\begin{enumerate}[(i)]
\item Section 1 presents a very general method that analytically calculates the error induced by any modifications of Maxwell's equations in the simulation domain. 
\item In sections 2 and 3, our analytical model is benchmarked against simulations. In particular, it is shown that our model can be used to deduce the growth rate of spurious truncation signals induced by the domain decomposition (section 2) as a function of various numerical parameters (Number of guard cells, order of Maxwell solver, subdomain sizes). It is also demonstrated that the new model enables very accurate predictions that are superior to previous models, of the efficiency of PMLs (section 3) with high-order stencils as a function of frequency and angle.
\end{enumerate}

\section{Model for errors induced by spatial modifications of Maxwell's equations in the simulation domain}
\label{sec:model}
This section presents a method for deriving analytically the error induced by the modifications of Maxwell's equations at an arbitrary number of nodes of the grid in the case of a plane and monochromatic wave.  The total error for an arbitrary waveform/beam is derived using plane wave decomposition and linearity of Maxwell's equations. 

When the simulation domain is uniform, the Von Neumann analysis \cite{TELLUSA8607} provides a very simple yet powerful means of studying the stability and growth of errors in the simulation domain due to the discretization of equations. However, as soon as the simulation domain has discontinuities at any point (e.g at domain boundaries, mesh irregularities or stencil variations) where the discretized Maxwell's equations are modified, it is no longer adapted. 

In the following we provide an alternative approach that provides accurate estimates of the total error $\zeta$ induced by the modifications of Maxwell's equations at arbitrary nodes of the simulation domain. In this paper, we consider the most common case of a Maxwell's field solver on a staggered grid \cite{Yee1966}. In appendix A, we give a formal definition of this scheme at order $p$ and verify that when $p\rightarrow\infty$, this solver converges to the pseudo-spectral solver \cite{PSTDLiu}, verifying that the new analysis model applies to pseudo-spectral solvers when $p\rightarrow\infty$. 

\subsection{Principle of the technique}

The principle of our analytical approach is progressively introduced in three steps by considering solutions of Maxwell's equations for the three different cases: 
\begin{enumerate}[(i)]
\item regular stencil in vacuum, 
\item regular stencil with an external monochromatic source point, 
\item irregular stencil in vacuum (case of interest). 
\end{enumerate}
\subsubsection{Regular stencil in vacuum}

Discretized Maxwell's equations in vacuum are written as: 
\begin{eqnarray}
\textbf{E}_{\textbf{r}}^{n+1}&=&\textbf{E}_{\textbf{r}}^{n}-c\delta t \nabla_p^* \times c \textbf{B}^{n+\frac{1}{2}}_{\textbf{s}}\label{eqEvac}\\
\textbf{B}_{\textbf{s}}^{n+\frac{1}{2}}&=&\textbf{B}_{\textbf{s}}^{n-\frac{1}{2}}-c\delta t \nabla_p^* \times c \textbf{E}^{n}_{\textbf{r}}\label{eqBvac}
\end{eqnarray}

where $n$ is the time step, \textbf{E} the electric field and \textbf{B} the magnetic field defined on staggered grids $\textbf{r}$ and $\textbf{s}$.  $\nabla_p^{*}$ is the discrete finite difference operator of order $p$ for the staggered scheme.

A plane wave of frequency $\omega$ will propagate with a wavevector $\textbf{k}$ given by the dispersion relation $k=f(\omega)$, where k is the norm of $\textbf{k}$. In 1D, this relation is: 
\begin{equation}
\sin\frac{\omega\delta t}{2}=\eta_x\sum_{l=1}^{p/2}C_{l}^{p}\sin(2l-1)k\frac{\delta x}{2}\label{disprel1D}
\end{equation}
where $\delta t$ and $\delta x$ are respectively the time step and mesh size, the coefficients $C_l^p$ are the stencil coefficients of the order $p$ staggered scheme solver (cf. Appendix A) and $\eta_x=c\delta t/\delta x$ is the Courant parameter.

\subsubsection{Regular stencil with an external monochromatic source point term}

We now investigate the discrete solutions of Maxwell's equations in 1D when an external monochromatic source point $\textbf{J}^n=\xi_0 e^{j_c\omega n \delta t}\delta(i)$, with $j_c^2=-1$, is introduced at position $i=0$ on the x-axis: 

\begin{eqnarray}
\textbf{E}_{\textbf{r}}^{n+1}&=&\textbf{E}_{\textbf{r}}^{n}-c\delta t \nabla_p^* \times c \textbf{B}^{n+\frac{1}{2}}_{\textbf{s}}\label{eqEvacext}+\textbf{J}^n\\
\textbf{B}_{\textbf{s}}^{n+\frac{1}{2}}&=&\textbf{B}_{\textbf{s}}^{n-\frac{1}{2}}-c\delta t \nabla_p^* \times c \textbf{E}^{n}_{\textbf{r}}\label{eqBvacext}
\end{eqnarray}

 As Maxwell's equations are linear, all the generated electromagnetic fields will also be monochromatic with frequency $\omega$ and Maxwell's equations (\ref{eqEvacext},\ref{eqBvacext}) can be written: 
 
\begin{eqnarray}
e_i\left(e^{j_c\omega\delta t/2}-e^{-j_c\omega\delta t/2}\right)&=&-\eta_x\sum_{l=1}^{p/2}C_l^p\left(b_{i+\frac{1}{2}+(l-1)}-b_{i+\frac{1}{2}-l}\right)+\xi_0 \delta(i)\label{eqE1Dvac}\\
b_{i+\frac{1}{2}}\left(e^{j_c\omega\delta t/2}-e^{-j_c\omega\delta t/2}\right)&=&-\eta_x\sum_{q=1}^{p/2}C_q^p\left(e_{i+q}-e_{i-(q-1)}\right)\label{eqB1Dvac}
\end{eqnarray}

where we used $E^n_i=e_ie^{j_c\omega n\delta t}$ and $B^{n+\frac{1}{2}}_{i+\frac{1}{2}}=b_{i+\frac{1}{2}}e^{j_c\omega n\delta t/2}$.  Replacing the expressions of $b_{i-l+\frac{1}{2}}$ and $b_{i+(l-1)+\frac{1}{2}}$ given by equation (\ref{eqB1Dvac}) in equation (\ref{eqE1Dvac}) yields the following equation for the electric field: 

\begin{equation}
4\sin^2\frac{\omega\delta t}{2}e_i+\eta_x^2\sum_{(l,q)=1}^{p/2}C_l^pC_l^q\left[e_{i+(q+l)-1}-(e_{i+(q-l)}+e_{i-(q-l)})+e_{i-(q+l)+1}\right]=-2j_c\sin\frac{\omega\delta t}{2}\xi_0\delta(i)
\end{equation}

which can be written in a more compact form: 

\begin{equation}
\sum_{l=-p+1}^{p-1}\kappa^{p}_le_{i+l}=-2j_c\sin\frac{\omega\delta t}{2}\xi_0\delta(i)\label{eqevacuum}
\end{equation}

where the coefficients $\kappa^{p}_l$ are functions of the stencil coefficients $C_l^{p}$, $\eta_x$ and $\omega\delta t$. These coefficients are symmetric and verify  $\kappa^{p}_l=\kappa^{p}_{-l}$. 

\begin{center}
\begin{figure}[!h]
\centering
\includegraphics[width=0.85\linewidth]{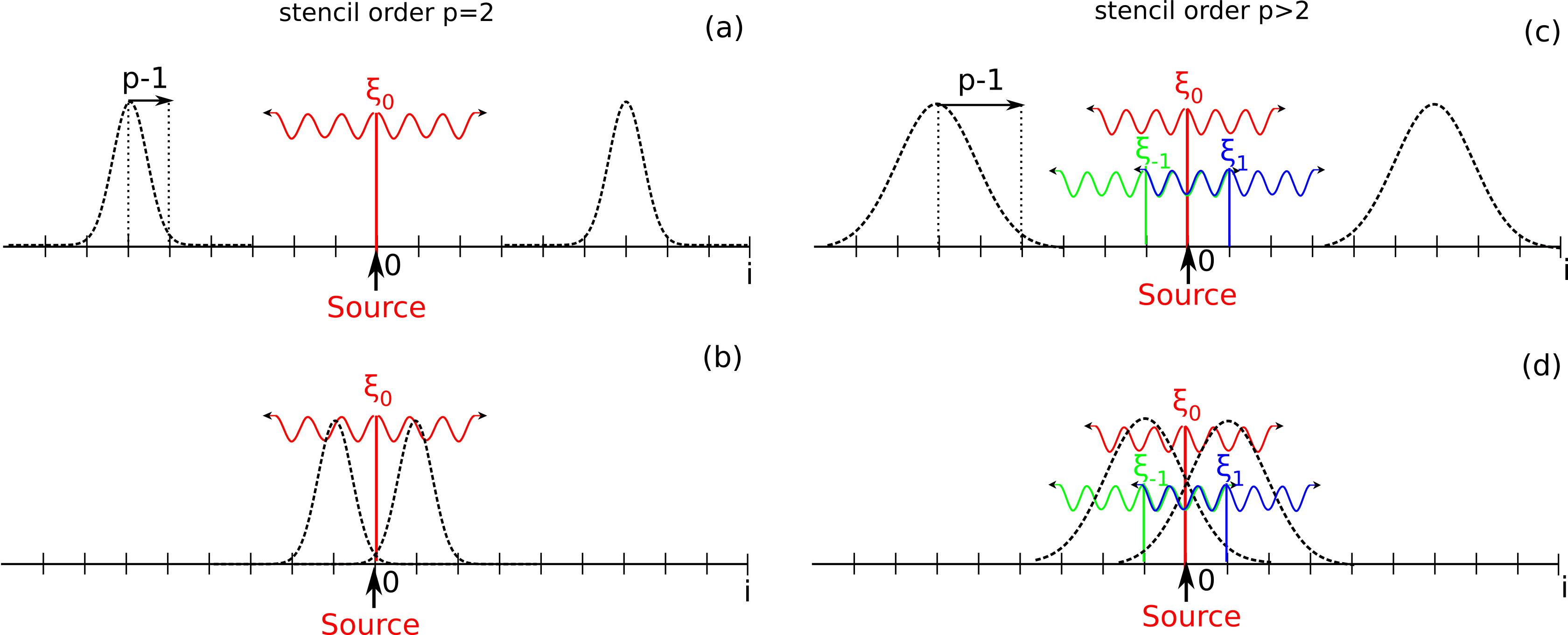}
\caption{Field emitted by a monochromatic external source point of amplitude $\xi_0$. In each panel, the black dotted line represents the envelope of the stencil with coefficients $\kappa_i$ used to compute $e_i$ from equation (\ref{eqevacuum}) at node $i$. The external monochromatic source is represented in red. Adjacent secondary sources, created by stencils crossing on each side of the primary source location for $p>2$, are represented in green and blue.  (a) Calculation of the field $e_i$ far from the external source point for order $p=2$. (b) Calculation of the field $e_i$ near the external source point for order $p=2$ (c) Calculation of the field $e_i$ far from the external source point for order $p>2$. (d) Calculation of field $e_i$ near the external source point for order $p>2$. }
\label{princ_exter_source}
\end{figure}
\end{center}

Calculating the electric field $e_{i}$ at node $i$ thus requires the values of the electric field at $2p-1$ adjacent nodes, as illustrated by the black dotted line on Fig. \ref{princ_exter_source}. The system of equations (\ref{eqevacuum}) written at all nodes $i$ of the domain is closed by considering that at positions $i_{+}>0$ and $i_{-}<0$ far from the external source point, equation (\ref{eqevacuum}) corresponds to a propagation equation in vacuum (cf. Fig \ref{princ_exter_source} (a) and (c)) and admits solutions of the form: 
\begin{equation}
e_{i_{\pm}}=\xi_v e^{\pm j_c i_{\pm} k \delta x}\label{solevac}
\end{equation}

where $k$ is solution of the dispersion relation $k=f(\omega)$ of the scheme in vacuum \ref{disprel1D} and $\xi_v$ the amplitude of the radiated wave in vacuum. Unrolling the system of equations (\ref{eqevacuum})  is equivalent to "retropropagating" the single waves given by equation (\ref{solevac}) from the far field to the initial source point $i=0$. 

For order $p=2$, equation (\ref{eqevacuum}) can be solved independently on subdomains $(i<0)$ and $(i>0)$ (cf. Fig \ref{princ_exter_source} (a) and (b)) and corresponds to a simple propagation in vacuum from both sides of the source point: 

\begin{equation}
e_{i}=\left \{\begin{array}{cc}
\xi_v e^{j_c k i\delta x}&i\leqslant 0\\
\xi_v e^{-j_c k i\delta x}&i\geqslant 0\\
\end{array}
\right.\label{solevacp2}
\end{equation}

with:
\begin{equation}
 \xi_v=\frac{\xi_0}{2\eta_x\cos\frac{k\delta x}{2}}
\end{equation}

For order $p>2$, the solution far from the source point (cf. Fig. \ref{princ_exter_source} (c)) is still given by equation (\ref{solevac}) and corresponds to a plane monochromatic wave propagating with wavevector $k$ given by the dispersion relation of the order $p$ scheme. 

However, near the external source point (i.e $-(p-1)< i <p-1$), equation (\ref{eqevacuum}) does not correspond to a propagation equation anymore and its solutions will not be the same as for order $2$, i.e single waves propagating in the backward (for i<0) or forward (for i>0) directions. 

Indeed, equations on the half domains $i<0$ and $i>0$ are now coupled due to a larger spatial extent of the high order stencil (cf Fig. \ref{princ_exter_source} (d)) and this coupling can be interpreted as the presence of secondary source points near the initial external source point. 

Considering the symmetry of the external current distribution with respect to $i=0$, the solutions $e_i=e_{-i}$  are also symmetric and equation (\ref{eqevacuum}) written at positions $-(p-1)< i_0<(p-1)$ ($i_{0}\neq0$) near the source point thus can be written:

\begin{equation}
\sum_{l=-(p-1)}^{(p-1)}\kappa^{p*}_l e_{i_0+l}=0\label{eqenear}
\end{equation}

where $\kappa_l^{p*}$ is different from its value $\kappa_l^{p}$ in vacuum. For the backward case $-(p-1)<i_{0}<0$, $\kappa_l^{p*}$ is: 

\begin{equation}
 \kappa_l^{p*}=\left \{\begin{array}{cc}
 \kappa_l^p    & -(p-1)<i+l<-i-(p-1)\\
 2\kappa_l^{p}& -i-(p-1)<i+l<0\\
0& 0<i+l<i+p-1\\
 \end{array}\right.
\end{equation}

While the solution given by equation (\ref{solevacp2}) satisfies the propagation equation (\ref{eqevacuum}) in vacuum for $i_0\leqslant -(p-1)$ or $i_0\geqslant(p-1)$, it does not satisfy anymore equation (\ref{eqenear}) as soon as $i_{0}\geqslant-(p-2)$ or $i_{0}\leqslant(p-2)$.

This modification of the electric field equation will thus change the value of the field $e_{i_0}$ at $i_0$ compared to a simple propagation in vacuum as given by equation (\ref{solevacp2}). This can be formally written as adding another monochromatic "secondary source" point in $i=i_{0}$ of amplitude $\xi_{i_0}$ that will modify the amplitude of the solution in vacuum.   

 


 
 As for the original external source point, this secondary source point also radiates an electromagnetic wave in $i<0$ and $i>0$ directions and thus contribute to the total electromagnetic field far from its position (cf Fig. \ref{princ_exter_source} (d)).  The total electric field thus has the following form: 
 
 \begin{equation}
e_{i}\propto\left \{\begin{array}{lr}
\sum_{-(p-1)<l<(p-1)}\xi_l  e^{j_c k (i-l)\delta x}&i\leqslant i_{-}=-(p-1)\\
\sum_{-(p-1)<i+l<i}\xi_l  e^{-j_c k (i-l)\delta x}+\sum_{i\leqslant i+l<p-1}\xi_l  e^{j_c k (i-l)\delta x}&-(p-2)\leqslant i\leqslant (p-2)\\
\sum_{-(p-1)<l<(p-1)}\xi_l  e^{-j_c k (i-l)\delta x}&i\geqslant i_{+}=(p-1)\\
\end{array}
\right.\label{etotp}
\end{equation}

Far from the external source, the total electric field given by equation (\ref{etotp}) can still be written as a single wave propagating forward (for $i\geqslant i_{+}$) or backward (for $i\leqslant i_{-}$) . This single wave results from the interference of multiple waves generated by the initial source point and by the $2(p-2)$ surrounding secondary sources.

 However, near the external source ($-(p-2)\leqslant i\leqslant (p-2)$), the form of the total electric field (\ref{etotp}) is more complex and is a linear combination of multiple waves propagating forwards and backwards (cf Fig. \ref{princ_exter_source} (d)).  

\subsubsection{Irregular stencil in vacuum}

In the following, we show that modifying Maxwell's equations at one node is equivalent to adding a "pseudo-source" term. In accordance with the result of the external source case described in the previous subsection, this induces the creation of additional adjacent secondary pseudo-sources around the modified node. 

Formally, local modifications of the stencil can be written as: 

\begin{eqnarray}
\textbf{E}_{\textbf{r}}^{n+1}&=&\alpha_r\textbf{E}_{\textbf{r}}^{n}-\beta_r c\delta t\nabla_p^{**} \times \textbf{B}^{n+\frac{1}{2}}_{\textbf{s}}\\
\textbf{B}_{\textbf{s}}^{n+\frac{1}{2}}&=&\alpha_s\textbf{B}_{\textbf{s}}^{n-\frac{1}{2}}-\beta_{s}c\delta t \nabla_p^{**} \times c \textbf{E}^{n}_{\textbf{r}}
\end{eqnarray}

where $\alpha$, $\beta$ and $\nabla_p^{**}$ are space varying coefficients and operator.

 Modifying the electric field equation at one node $\textbf{r}_0=(r_{0,x},r_{0,y},r_{0,z})$ only of the simulation domain can thus  be written  as: 
\begin{equation}
\textbf{E}_{\textbf{r}}^{n+1}=\textbf{E}_{\textbf{r}}^{n}-c\delta t\nabla_p^{*} \times \textbf{B}^{n+\frac{1}{2}}_{\textbf{s}}+\textbf{J}^{**}
\end{equation}
which is the same equation as in vacuum but with an additional time-dependent pseudo-source term $\textbf{J}^{**}$: 
\begin{equation}
\textbf{J}^{**}=\left[(-1+\alpha_r)\textbf{E}_{\textbf{r}}^{n}-\beta_r c\delta t(\nabla_p^{**}-\nabla_p^{*})\times \textbf{B}^{n+\frac{1}{2}}_{\textbf{s}}\right]\delta(\textbf{r}-\textbf{r}_0)
\end{equation}

For a monochromatic driving wave of frequency $\omega$ propagating with wavevector $\textbf{k}$ given by the dispersion relation in vacuum (far from the modification), the modification of Maxwell's equations at $\textbf{r}_0$ will thus generate a monochromatic pseudo-current $\textbf{J}^{**}$ at $\textbf{r}=\textbf{r}_0$ (cf. Fig. \ref{principlemodel} (a)): 
\begin{equation}
\textbf{J}^{**}= \xi_0 e^{j_c \omega n \delta t}\delta(\textbf{r}-\textbf{r}_0)\label{eqJ**}
\end{equation} 
Similarly to the previous case of an external point source term at order $p>2$, this "pseudo-source" term will induce the creation of $(2p-2)$ monochromatic secondary source terms of amplitudes $(\xi_{\textbf{r}_0+l})_{-(p-2)\leqslant l\leqslant (p-2)}$ in each dimension around the modified node $\textbf{r}_0$, that will also radiate and contribute to the total electric field. These coefficients will be further called re-emission coefficients. 


A similar reasoning on $\textbf{B}$ shows that $2(p-2)$ time-dependent secondary pseudo-source terms are produced around modified internodes. 

\subsubsection{New features added by the model}

\begin{center}
\begin{figure}[!h]
\centering
\includegraphics[width=0.65\linewidth]{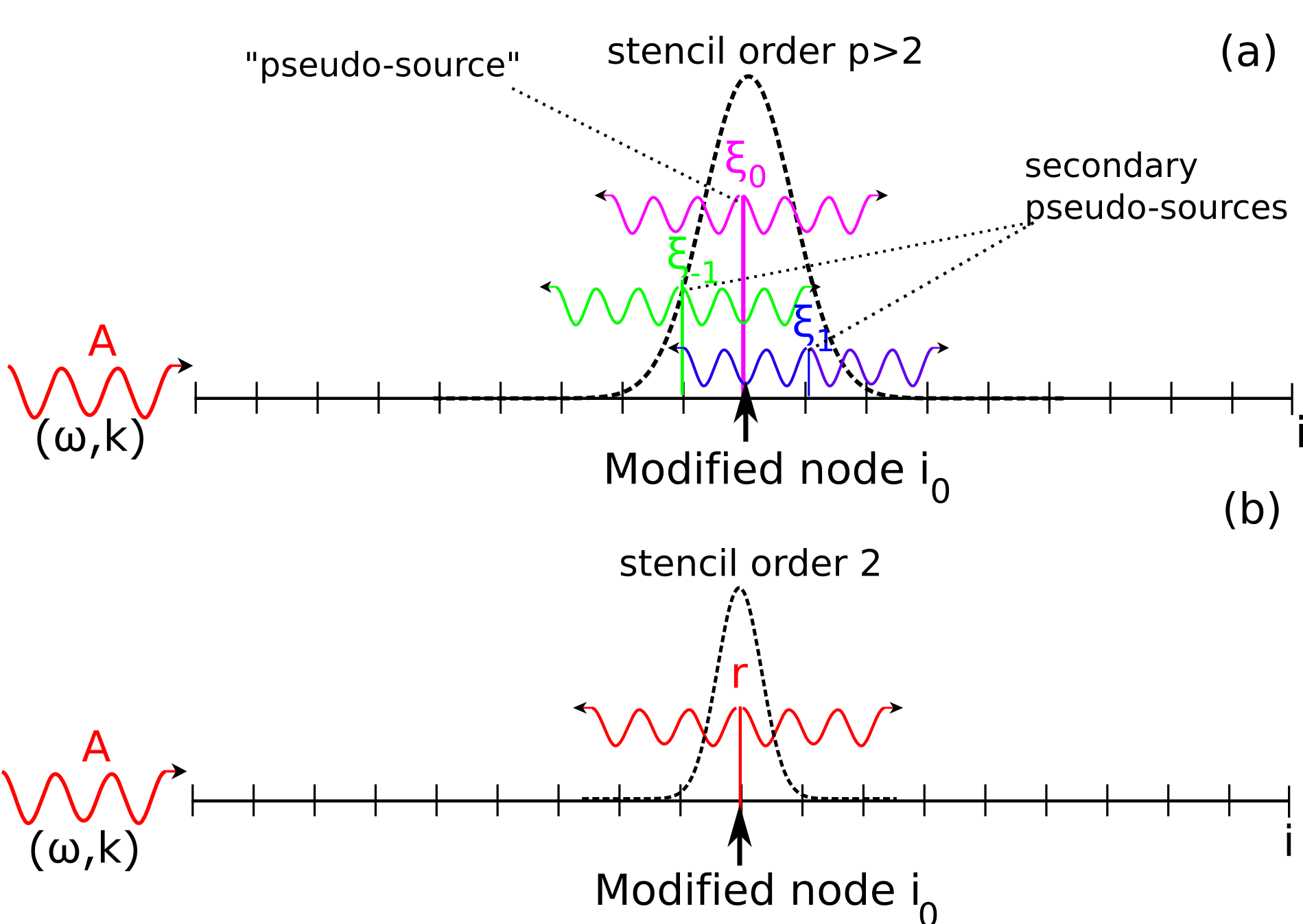}
\caption{Principle of the model. (a) The new model takes into account the fact that at order $p$, several nodes $l$ around the modified node act as monochromatic "secondary pseudo-sources" of error with amplitudes $\xi_l$, frequency $\omega$ (equals to the incident wave frequency) and wave vector $k$ given by the dispersion relation in vacuum. (b) Former model considered that only the modified node is acting as a monochromatic source of error with amplitude $r$. }
\label{principlemodel}
\end{figure}
\end{center}

Former techniques \cite{Vay2002, Lee2015} (see Fig. \ref{principlemodel} (b)) rather made the assumption that modifying Maxwell's equations at one $i=i_0$ node would lead to the reflection of the driving wave at this particular node with a reflection coefficient $r$ and transmission coefficient $1-r$. This would be equivalent to adding only one pseudo source at position $i_0$ with a re-emission coefficient $\xi_0=-r$ that radiates in both directions $i<i_0$ and $i>i_0$. While this is adequate for the analysis of second order FDTD schemes for which the former model was developed initially \cite{Vay2002}, the assumption of a single reemission pseudo-source leads to discrepancy between modeling and simulations for the coefficient $r$ at orders $p>2$ \cite{Lee2015}. 

This inaccurate estimate becomes highly detrimental when looking for instance at stencil truncations with the domain decomposition technique at very high orders $p$ \cite{Vay2013}. As explained in the above section, the discrepancy is due to the fact that for Maxwell solvers of orders $p>2$, the  $2(p-2)$-adjacent nodes near $i_0$ will also be affected by modification of the fields at $i_0$. As a consequence, they will also act as secondary pseudo-sources that radiate monochromatic waves. 

In the following section a method is introduced for computing the reemission coefficients $(\xi_{\textbf{r}_0+l})_{-(p-2)\leqslant l\leqslant (p-2)}$ of secondary pseudo-sources and deduce the total amplitude $\zeta$ of the error generated by their interference. We will call this new method the "p-sources" or "multi-sources" model as opposed to the "one-source" models developed earlier. 
\newpage
\subsection{1D Model: Detailed analytical calculation of the coefficients $\xi_l$}

\subsubsection{Initial assumptions}

 \begin{center}
\begin{figure}[h]
\centering
\includegraphics[width=0.79\linewidth]{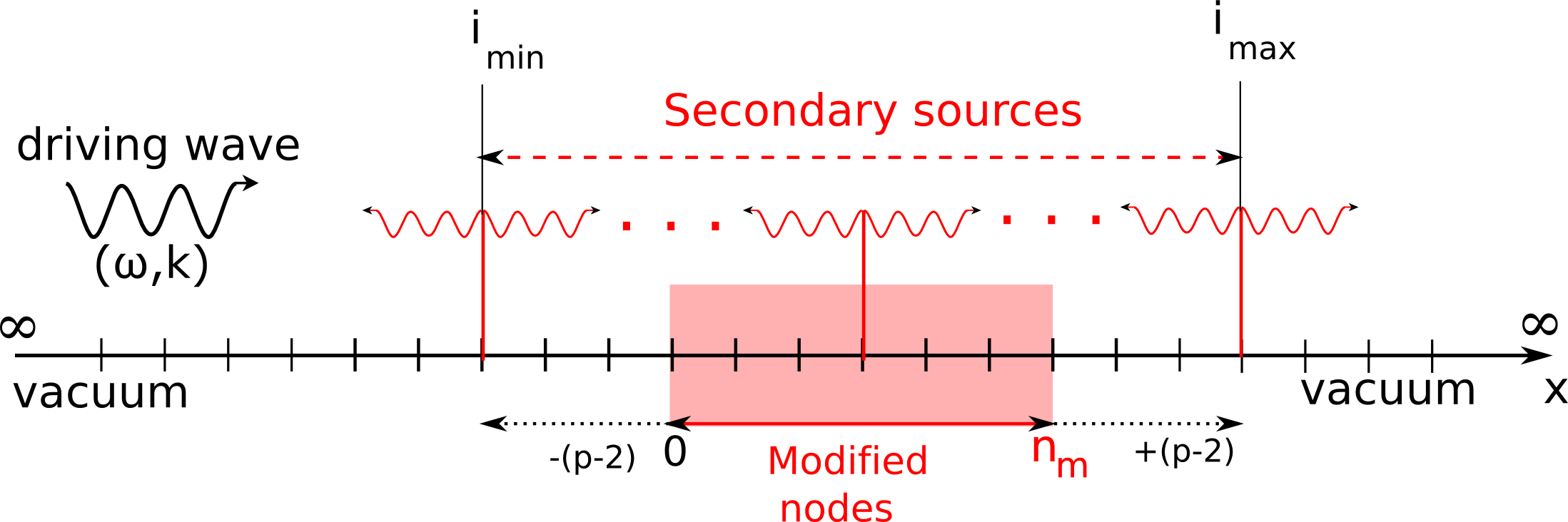}
\caption{Configuration of the problem in a 1D geometry. Maxwell's equations are modified at $n_m$ nodes of the grid. As shown in the previous subsection, the $2(p-2)$ adjacent nodes at left and right of the modified nodes also act as "secondary pseudo-sources". }
\label{pbconfiguration}
\end{figure}
\end{center}

 The problem layout is detailed in Fig. \ref{pbconfiguration}. For the sake of simplicity, the method is first detailed in a 1D geometry for a staggered field solver of order $p$, and is generalized for a 3D geometry in the following subsection. 
 
 We assume a linearly p-polarized $(E_{d},B_{d})$ driving wave of frequency $\omega$ propagating along the x-axis (Electric field in the plane (x,y) and Magnetic field in the plane (x,z)). 
%

%
We consider an infinite domain along $x$ and assume that Maxwell's equations are modified at $n_m+1$ nodes and $n_m$ internodes of the grid, from $i=0$ to $i=n_m$. As a consequence, nodes $i$ and internodes $i+\frac{1}{2}$ ranging from $i_{min}=-(p-2)$ to $i_{max}=n_m+(p-2)$ will act as secondary pseudo-sources of amplitudes $(\xi_{i+l})_{-(p-2)\leqslant l\leqslant p-2}$ and $(\xi_{i+\frac{1}{2}l})_{-(p-2)\leqslant l\leqslant (p-2)}$. Each one of this individual source will emit electromagnetic radiations in both $(x<0)$ and $(x>0)$ directions. 

\subsubsection{Expression for magnetic $B_z$ and electric $E_y$ fields}

\begin{center}
\begin{figure}[h!]
\centering
\includegraphics[width=0.99\linewidth]{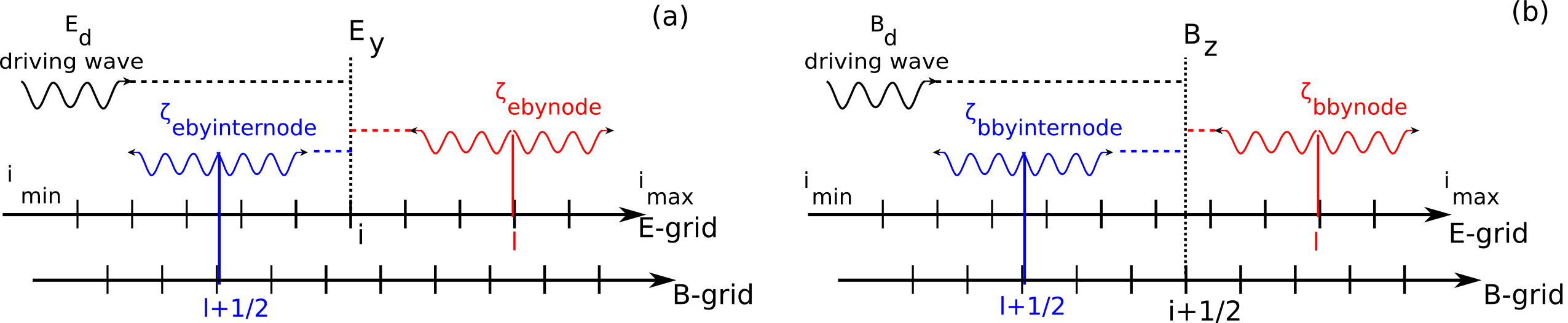}
\caption{Contribution of adjacent sources and driving wave to the total electric and magnetic fields $(E_y,B_z)$. (a) The expression of the electric field $E_{y,i}$ at node $i$ is a combination of electric field $E_{d,i}$ at position $i$ of the driving plane wave, the electric field $\zeta_{ebynode,i,l}$ at position $i$ emitted by the secondary source located at position $l$ ($i_{min}<l<i_{max}$, the electric field $\zeta_{ebyinternode,i,l+\frac{1}{2}}$ at position $i$ emitted by the secondary source located at internode $l+\frac{1}{2}$ ($i_{min}<l+\frac{1}{2}<i_{max}$). (b) The expression of the magnetic field $B_{z,i}$ at node $i$ is a combination of magnetic field $B_{d,i}$ at position $i$ of the driving plane wave, the magnetic field  $\zeta_{bbynode,i,l}$ at position $i$ emitted by the secondary source located at position $l$ ($i_{min}<l<i_{max}$, the magnetic field $\zeta_{bbyinternode,i,l+\frac{1}{2}}$ at position $i$ emitted by the secondary source located at internode $l+\frac{1}{2}$ ($i_{min}<l+\frac{1}{2}<i_{max}$). }
\label{field:form}
\end{figure}
\end{center}

From the analysis in the preceding subsections, we can infer a form for the total electric and magnetic fields ($E_{y,i}^n$,$B_{z,i}^n$) at position $i$ and time $n$ which are the combination of three components for the electric field $E_y$ (see Fig. \ref{field:form} (a)): 

\begin{eqnarray}
E_{y,i}^n&=&E_{d,i}^n+\sum_{l=i_{min}}^{i_{max}}\zeta_{ebynode,i,l}^{n}+\sum_{l=i_{min}+\frac{1}{2}}^{i_{max}-\frac{1}{2}}\zeta_{ebyinternode,i,l+\frac{1}{2}}^{n}\label{eqForm1D:Ey}\\
B_{z,i}^n&=&B_{d,i}^n+\sum_{l=i_{min}}^{i_{max}}\zeta_{bbynode,i,l}^{n}+\sum_{l=i_{min}+\frac{1}{2}}^{i_{max}-\frac{1}{2}}\zeta_{bbyinternode,i,l+\frac{1}{2}}^{n}\label{eqForm1D:Bz}
\end{eqnarray}

where the individual components are defined below:
\begin{itemize}
\item the electric field $E_{d,i}^{n}$ at position $i$ corresponding to the electric field of the driving plane wave: 
\begin{equation}
E_{d,i}^{n}=e^{-j_cki\delta x +\omega n \delta t}
\end{equation}
\item the electric field $\zeta_{ebynode,i,l}^{n}$ at position $i$ emitted by the secondary source at position $l$ ($i_{min}<l<i_{max}$):
\begin{equation}
\zeta_{ebynode,i,l}^{n}=\left \{\begin{array}{cc}
\xi_l e^{j_c k i\delta x+\omega n \delta t}& i\leqslant l\\
\xi_l e^{-j_c k i\delta x+\omega n \delta t}& i\geqslant l\\
\end{array}\right.
\end{equation}
\item the electric field $\zeta_{ebyinternode,i,l+\frac{1}{2}}^{n}$ at position $i$ emitted by the secondary source at internode $l+\frac{1}{2}$ ($i_{min}<l+\frac{1}{2}<i_{max}$):
\begin{equation}
\zeta_{ebyinternode,i,l+\frac{1}{2}}^{n}=\left \{\begin{array}{cc}
-\xi_{l+\frac{1}{2}} e^{j_c k i\delta x+\omega n \delta t}& i\leqslant l\\
\xi_{l+\frac{1}{2}} e^{-j_c k i\delta x+\omega n \delta t}& i\geqslant l\\
\end{array}\right.
\end{equation}
\end{itemize}
and similarly three components for the magnetic field $B_z$  (see Fig. \ref{field:form} (b)):

\begin{itemize}
\item the magnetic field $B_{d,i}^{n}$ at position $i$ corresponding to the electric field of the driving plane wave (see Fig. \ref{field:form} (b)): 
\begin{equation}
B_{d,i}^{n}=e^{-j_cki\delta x +\omega n \delta t}
\end{equation}
\item the magnetic field $\zeta_{bbyinternode,i,l+\frac{1}{2}}^{n}$ at position $i$ emitted by the secondary source at internode $l+\frac{1}{2}$ ($i_{min}<l+\frac{1}{2}<i_{max}$):
\begin{equation}
\zeta_{bbyinternode,i,l+\frac{1}{2}}^{n}=\left \{\begin{array}{cc}
\xi_{l+\frac{1}{2}} e^{j_c k i\delta x+\omega n \delta t}& i\leqslant l\\
\xi_{l+\frac{1}{2}} e^{-j_c k i\delta x+\omega n \delta t}& i\geqslant l\\
\end{array}\right.
\end{equation}
\item the magnetic field $\zeta_{bbynode,i,l}^{n}$ at position $i$ emitted by the secondary source at internode $l$ ($i_{min}<l<i_{max}$):
\begin{equation}
\zeta_{bbynode,i,l}^{n}=\left \{\begin{array}{cc}
-\xi_l e^{j_c k i\delta x+\omega n \delta t}& i\leqslant l\\
\xi_l e^{-j_c k i\delta x+\omega n \delta t}& i\geqslant l\\
\end{array}\right.
\end{equation}
\end{itemize}

\subsubsection{Calculation of the total error $\zeta$ in the general case}

For a staggered field solver of order $p$, Maxwell's discrete equations for the electric and magnetic fields can be written as: 
\begin{equation}
\begin{array}{lcl}
E_{y,i}^{n+1}&=&\alpha_i E_{y,i}^{n}\\
&-&\beta_{i} \left[\sum_{l=1}^{\frac{p}{2}}\Gamma_{i,l}^{p} cB_{z,i+\frac{1}{2}+(l-1)}^{n+1/2}- \sum_{l=1}^{\frac{p}{2}}\psi_{i,l}^{p}cB_{z,i+\frac{1}{2}-l}^{n+\frac{1}{2}}\right]\\
\end{array}
\label{eqEi}
\end{equation}

and:

\begin{equation}
\begin{array}{lcl}
 cB_{z,i+\frac{1}{2}}^{n+\frac{1}{2}}&=&\alpha_{i+\frac{1}{2}}cB_{z,i+\frac{1}{2}}^{n-\frac{1}{2}}\\
 &-&\beta_{i+\frac{1}{2}}  \left[\sum_{q=1}^{\frac{p}{2}}\Gamma_{i+\frac{1}{2},q}^{p} E_{y,i+q}^{n}-\sum_{q=1}^{\frac{p}{2}}\psi_{i+\frac{1}{2},q}^{p}E_{y,i-(q-1)}^{n}\right]\end{array}\label{eqBi}
\end{equation}

with $\alpha_i$, $\beta_{i}$, $\Gamma_{i,l}^{p}$ and  $\psi_{i,l}^{p}$ coefficients that may vary with the position on the grid $i$. In vacuum, we have $\alpha_i=1$, $\beta_i=-c\delta t/\delta x$ and $\Gamma_{i,l}^{p}=\psi_{i,l}^{p}=C_{l}^{p}$ for all positions $i$ on the grid. 

For the following, let us assume that the coefficients $\alpha_i$, $\beta_{i}$, $\Gamma_{i,l}^{p}$ and  $\psi_{i,l}^{p}$ are modified between nodes $i=0$ and $i=n_m$. Under these assumptions, the unknowns of the system of equations with varying coefficients are the $N=2(i_{max}-i_{min})+1$ re-emission coefficients coefficients $(\xi_{l})_{i_{min}\leqslant l \leqslant i_{max} }$ for nodes and $(\xi_{l+\frac{1}{2}})_{i_{min}\leqslant l+\frac{1}{2}\leqslant i_{max} }$ for internodes. Finding them requires the $N_E=(i_{max}-i_{min})+1$ equations (\ref{eqEi})  for the electric field $E_y$ written at nodes ${i_{min}\leqslant i\leqslant i_{max}}$ and the $N_B=(i_{max}-i_{min})$ equations  (\ref{eqBi}) for the magnetic field $B_z$ written at internodes ${i_{min}\leqslant i+\frac{1}{2}\leqslant i_{max}}$. Note that the closure of the system is achieved thanks to the finite number of reemission sources, as it is assumed that there is no additional sources beyond  $i<i_{min}$ and $i>i_{max}$ (vacuum).

In vacuum, it can be shown that the system to be solved can be expressed in the following matrix form: 

\begin{equation}
A_vX=0
\label{eqvac}
\end{equation}

with: 

\begin{equation}
A_v=
\begin{bmatrix}
a_{0}&a_{1}&\cdots& a_{p/2}&.. &0 \\
a_{-1}&a_{0}&\cdots&\cdots&\ddots&\vdots \\
\vdots & \ddots & \ddots & \ddots&\ddots&a_{p/2}\\
a_{-p/2} & \ddots& \ddots & \ddots&\ddots&\vdots\\
\vdots & \ddots & \ddots & \ddots&\ddots&a_1\\
0&\cdots&a_{-p/2}&\cdots&a_{-1} &a_{0}\\
\end{bmatrix}
\end{equation}

a band Matrix of size $N\times N$ and bandwidth $p+1$ (total width of the stencil). X is a vector of size $N\times1$ containing the unknowns of the problem: 

\begin{equation}
X=\begin{bmatrix}
\xi_{i_{min}} \\
\vdots\\
\xi_{i_{max}} \\
\end{bmatrix}
\end{equation}

 Note that equation (\ref{eqvac}) trivially yields $X=0$ (providing that $\det A\neq0$) in vacuum i.e $(\xi_{l+\frac{1}{2}})_{i_{min}\leqslant l+\frac{1}{2}\leqslant i_{max} }=0$ and $(\xi_{l})_{i_{min}\leqslant l \leqslant i_{max} }$=0.  If Maxwell's equations are modified at $n_m$ nodes and $n_m$ internodes, the system takes the form: 

\begin{equation}
A_m X=B
\label{eqmod}
\end{equation}

with:
\begin{equation}
A_m=A_v+
\begin{bmatrix}
0&\cdots&\cdots& \cdots&\cdots&0 \\
\vdots&\vdots&\vdots&\vdots&\vdots&\vdots \\
d_{0} & \cdots& \cdots & \cdots&\cdots&d_{0}\\
\vdots&\vdots&\vdots&\vdots&\vdots&\vdots \\
d_{2n_m-1} & \cdots& \cdots & \cdots&\cdots&d_{2n_m-1}\\
\vdots&\vdots&\vdots&\vdots&\vdots&\vdots \\
0&\cdots&\cdots& \cdots&\cdots&0 \\
\end{bmatrix}
\end{equation}

and, 

\begin{equation}
B=\begin{bmatrix}
0 \\
\vdots\\
b_{0}\\
\vdots\\
b_{2n_m-1} \\
\vdots\\
0\\
\end{bmatrix}
\end{equation}

The linear system of equation (\ref{eqmod}) can be easily solved analytically using Cramer's rule. The total re-emission coefficient $\zeta$  for the electric field in the $x>0$ direction is then simply given by: 

\begin{equation}
\zeta=\sum_{l=i_{min}}^{i_{max}}\xi_l e^{j k (l-i)\delta x}-\sum_{l=i_{min}}^{i_{max}}\xi_{l+1/2} e^{j k (l+1/2-i)\delta x}
\label{eqr}
\end{equation}

\subsubsection{Analytical resolution of the system}

Analytically, Cholesky LU decomposition \cite{ALCholesky1910} that runs in $O(N^3)$ may be used. However, as the system to be solved is sparse, iterative methods for sparse matrixes that scale in $O(NLogN)$ are used instead. For low order systems, the resolution can be done "by-hand" but requires advanced symbolic calculations at large orders. Results discussed in the remainder of this paper were obtained using Mathematica \cite{Mathematica8} and the LinearSolve function that has accelerating techniques for sparse matrixes, allowing fast calculation of numerical solutions for the coefficients $\xi_l$. 

\newpage
\subsection{Extension of the model to 3D}

\subsubsection{Initial assumptions}

We consider a 3D geometry and the case of a driving p-polarized $(E_{x},E_{y},E_{z},B_{x},B_{z})$ plane wave of frequency $\omega$ propagating with an angle $\theta$ in the $(\textbf{x},\textbf{y})$ plane and an angle $\phi$ in the $(\textbf{x},\textbf{z})$ plane. In vacuum the wave vector $\textbf{k}=(k_x,k_y,k_z)=(k \cos \theta\cos\phi,k\sin\theta,k\cos\theta\sin\phi)$ of such a wave is simply given by the 3D dispersion relation of the staggered scheme.

In the case of  p-polarized driving wave the $E$ field is in the plane of incidence $(\textbf{x},\textbf{y})$ and the $B$ field is orthogonal to this plane. It can be easily shown that equations are of the same form with a s-polarized driving wave where the $B$ field is in the plane of incidence and the $E$ field orthogonal to the plane of incidence. Results from arbitrary polarization can then be easily deduced by noticing that any polarization is the superposition of these two orthogonal polarization states, justifying the restriction of the analysis to the p-polarized case only without loss of generality. 

\begin{center}
\begin{figure}[h]
\centering
\includegraphics[width=0.69\linewidth]{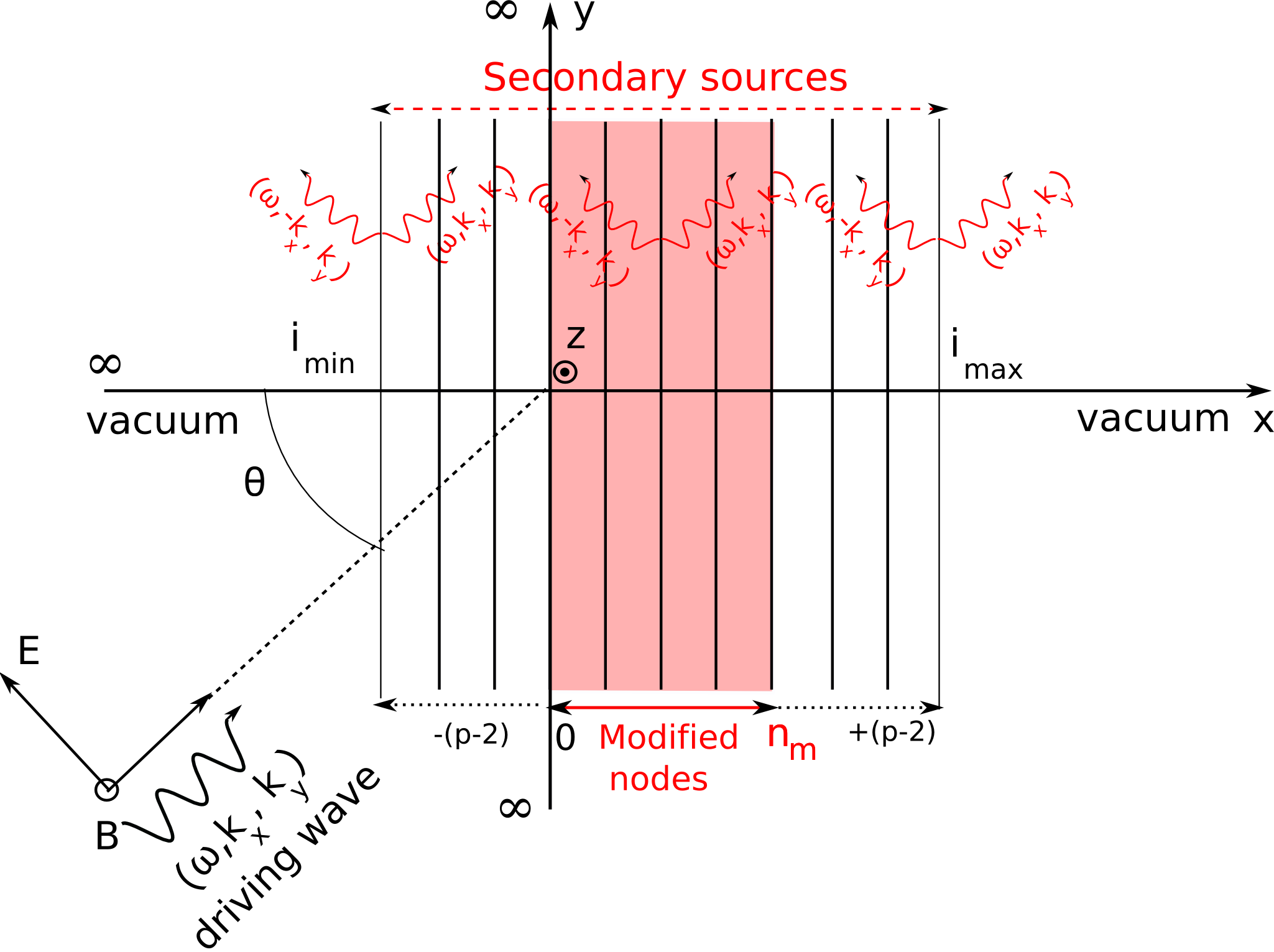}
\caption{Problem configuration in the 3D case ($\phi=0^o$, $k_z=0$). On this illustration, the driving wave is obliquely incident with an angle $\theta$ and $\phi=0^o$ on a set of $n_{m}$ modified nodes. Modified nodes act as secondary source planes that radiate in $x<0$ and $x>0$ directions.}
\label{model:config2D}
\end{figure}
\end{center}

The configuration of the 3D problem is sketched on Fig. \ref{model:config2D} in the case $\phi=0^o$ and $\theta\neq 0$. We consider an infinite domain along $x$, $y$ and $z$ axes and assume that Maxwell's equations are modified along direction $x$ only (translational invariance along $y$ and $z$), at $n_m$ nodes and $n_m$ internodes of the grid, from $i=0$ to $i=n_m$. In this case, nodes and internodes from $i_{min}=-(p-2)$ to $i_{max}=n_m+(p-2)$ will act as a secondary pseudo-source planes of amplitudes $(\xi_{i+l})_{-(p-2)\leqslant l\leqslant (p-2)}$ and $(\xi_{i+\frac{1}{2}+l})_{-(p-2)\leqslant l\leqslant (p-2)}$. 

As there is no stencil variations along $y$ and $z$, the spatial phase shift induced by stencil modifications will not depend on $j$ and $k$. This implies that there is no spread in propagation direction of the secondary sources. Consequently, each one of this individual source plane will emit plane waves in $x>0$ and $x<0$ directions with angles $\theta$ and $\pi-\theta$ with regard to the $x$ axis (see Fig. \ref{model:config2D}).

\subsubsection{General form of modified Maxwell's equations in $3D$}

In the 3D case, Maxwell's equations for the electromagnetic fields $(E_{x},E_{y},E_{z},B_{x},B_{z})$ have the following general form: 

\begin{flalign}\label{eqExijk}
&\begin{array}{lcl}
E_{x,i+\frac{1}{2},j,k}^{n+1}&=& \alpha_{x,i+\frac{1}{2}}E_{x,i+\frac{1}{2},j,k}^{n}\\
&+& \beta_{x,i+\frac{1}{2}}\sum_{l=1}^{\frac{p}{2}}C_{l}^{p}\left[ cB_{z,i+\frac{1}{2},j+\frac{1}{2}+(l-1),k}^{n+1/2}-cB_{z,i+\frac{1}{2},j+\frac{1}{2}-l,k}^{n+\frac{1}{2}}\right]\\
\end{array}&
\end{flalign}

\begin{flalign}\label{eqEyijk}
&\begin{array}{lcl}
E_{y,i,j+\frac{1}{2},k}^{n+1}&=&\alpha_{y,i} E_{y,i,j+\frac{1}{2},k}^{n}\\
&+& \beta_{y,i}\sum_{l=1}^{\frac{p}{2}}C_{l}^{p}\left[ cB_{x,i,j+\frac{1}{2},k+\frac{1}{2}+(l-1)}^{n+1/2}-cB_{x,i,j+\frac{1}{2},k+\frac{1}{2}-l}^{n+\frac{1}{2}}\right]\\
&-&\gamma_{y,i} \left[\sum_{l=1}^{\frac{p}{2}}\Gamma_{i,l}^{p} cB_{z,i+\frac{1}{2}+(l-1),j+\frac{1}{2},k}^{n+1/2}- \sum_{l=1}^{\frac{p}{2}}\psi_{i,l}^{p}cB_{z,i+\frac{1}{2}-l,j+\frac{1}{2},k}^{n+\frac{1}{2}}\right]\\
\end{array}&
\end{flalign}

\begin{flalign}\label{eqEzijk}
&\begin{array}{lcl}
E_{z,i,j,k+\frac{1}{2}}^{n+1}&=&\alpha_{z,i}E_{z,i,j,k+\frac{1}{2}}^{n}\\
&-& \beta_{z,i}\sum_{l=1}^{\frac{p}{2}}C_{l}^{p}\left[ cB_{x,i,j+\frac{1}{2}+(l-1),k+\frac{1}{2}}^{n+1/2}-cB_{x,i,j+\frac{1}{2}-l,k+\frac{1}{2}}^{n+\frac{1}{2}}\right]\\
\end{array}&
\end{flalign}

\begin{flalign}
&\begin{array}{lcl}
 cB_{x,i,j+\frac{1}{2},k+\frac{1}{2}}^{n+\frac{1}{2}}&=& \alpha^{*}_{x,i}cB_{x,i,j+\frac{1}{2},k+\frac{1}{2}}^{n-\frac{1}{2}}\\
 &-&\beta^{*}_{x,i}\sum_{l=1}^{\frac{p}{2}}C_{l}^{p}\left[ E_{z,i,j+l,k+\frac{1}{2}}^{n}-E_{z,i,j-(l-1),k+\frac{1}{2}}^{n}\right] \\
 &+&\gamma^{*}_{x,i}\sum_{l=1}^{\frac{p}{2}}C_{l}^{p}\left[ E_{y,i,j+\frac{1}{2},k+l}^{n}-E_{y,i,j+\frac{1}{2},k-(l-1)}^{n}\right]\\ \end{array}&\label{eqBxijk}
\end{flalign}


\begin{flalign}
&\begin{array}{lcl}
 cB_{z,i+\frac{1}{2},j+\frac{1}{2},k}^{n+\frac{1}{2}}&=&\alpha^{*}_{z,i+\frac{1}{2}}cB_{z,i+\frac{1}{2},j+\frac{1}{2},k}^{n-\frac{1}{2}}\\
 &-&\beta^{*}_{z,i+\frac{1}{2}} \left[\sum_{l=1}^{\frac{p}{2}}\Gamma_{i+\frac{1}{2},l}^{p} E_{y,i+l,j+\frac{1}{2},k}^{n}-\sum_{l=1}^{\frac{p}{2}}\psi_{i+\frac{1}{2},l}^{p}E_{y,i-(l-1),j+\frac{1}{2},k}^{n}\right]\\
 &+&\gamma^{*}_{z,i+\frac{1}{2}} \sum_{l=1}^{\frac{p}{2}}C_{l}^{p}\left[ E_{x,i+\frac{1}{2},j+l,k}^{n}-E_{x,i+\frac{1}{2},j-(l-1),k}^{n}\right] \end{array}&\label{eqBzijk}
\end{flalign}

$\alpha_{x,y,z}$, $\beta_{x,y,z}$, $\alpha^{*}_{x,z}$, $\beta^{*}_{x,z}$,$\gamma_{y,z}$, $\gamma^{*}_{x,z}$ are coefficients that may vary with index $i$ (not $j$ or $k$ in this configuration). In the above equations the discrete $\nabla_x$ operator is modified by taking different stencil coefficients $\Gamma_{i,l}^{p}$ and  $\psi_{i,l}^{p}$  that may also vary with the position on the grid $i$.  $\nabla_y$ and  $\nabla_z$ are not modified along $j$ and $k$. In vacuum, the coefficients are given by: 
\begin{eqnarray}
\alpha_{x,i}&=&\alpha^{*}_{x,i}=\alpha_{y,i}=\alpha_{z,i}=\alpha^{*}_{z,i}=1\\
\beta_{z,i}&=&\beta^{*}_{z,i}=\gamma_{y,i}=\gamma^{*}_{y,i}=\eta_{x}\\
\beta_{x,i}&=&\beta^{*}_{x,i}=\gamma^{*}_{z,i}=\eta_{y}\\
\gamma^{*}_{x,i}&=&\beta_{y,i}=\eta_{z}\\
\end{eqnarray}

and

\begin{equation}
 \Gamma_{i,l}^{p}=\psi_{i,l}^{p}=C_{l}^{p}
 \end{equation}
 In our configuration, Maxwell's equations between nodes $i=0$ and $i=n_m$ are modified so that coefficients $\alpha$, $\beta$, $\gamma$, $\alpha^{*}$, $\beta^{*}$, $\gamma^{*}$, $\Gamma$ and  $\psi$  may change between these positions. 
 
 \subsubsection{System of equations to solve in the 3D case}

Due to translational invariance along $j$ and $k$, equations (\ref{eqExijk},\ref{eqEzijk},\ref{eqBxijk}) can be written as: 

\begin{equation}
E_{x,i+\frac{1}{2},j,k}^{n}=\beta_{x,i+1/2}\frac{\sum_{l=1}^{\frac{p}{2}}C_{l}^{p}\left(e^{-j_ck_y(l-1)\delta y}-e^{j_ck_y l \delta y}\right) }{e^{j_c\omega \delta t}-\alpha_{x,i+1/2}}\times cB_{z,i+\frac{1}{2},j+\frac{1}{2},k}^{n+1/2}
\label{eqExvsBz}
\end{equation}

\begin{equation}
E_{z,i,j,k+\frac{1}{2}}^{n}=-\beta_{z,i+1/2}\frac{\sum_{l=1}^{\frac{p}{2}}C_{l}^{p}\left(e^{-j_ck_y(l-1)\delta y}-e^{j_ck_y l \delta y}\right) }{e^{j_c\omega \delta t}-\alpha_{y,i+1/2}}\times cB_{x,i,j+\frac{1}{2},k+\frac{1}{2}}^{n+1/2}
\label{eqEzvsBx}
\end{equation}

\begin{equation}
\begin{array}{lcl}
cB_{x,i,j+\frac{1}{2},k+\frac{1}{2}}^{n+\frac{1}{2}}&=&-\beta^{*}_{x,i}e^{j_c\omega \delta t}\frac{\sum_{l=1}^{\frac{p}{2}}C_{l}^{p}\left(e^{-j_ck_y l\delta y}-e^{j_ck_y (l-1) \delta y}\right) }{e^{j_c\omega \delta t}-\alpha^{*}_{x,i}}\times E_{z,i,j,k+\frac{1}{2}}^{n}\\
&+&\gamma^{*}_{x,i}e^{j_c\omega \delta t}\frac{\sum_{l=1}^{\frac{p}{2}}C_{l}^{p}\left(e^{-j_ck_z l\delta z}-e^{j_ck_z (l-1) \delta z}\right) }{e^{j_c\omega \delta t}-\alpha^{*}_{x,i}}\times E_{y,i,j+\frac{1}{2},k}^{n}\\
 \end{array}
\label{eqBxvsExEy}
\end{equation}

By expressing $E_x$ and $E_z$ as a function of $E_y$ and $B_z$ using the above equations (\ref{eqExvsBz}), (\ref{eqEzvsBx}) and (\ref{eqBxvsExEy}), the system of equations (\ref{eqExijk}-\ref{eqBzijk}) can be re-written into a system of two equations only on $E_{y}$ and $B_{z}$: 

\begin{flalign}\label{eqEymodi}
&\begin{array}{lcl}
E_{y,i}^{n+1}&=&a_i E_{y,i}^{n}\\
&-&\gamma_{y,i} \left[\sum_{l=1}^{\frac{p}{2}}\Gamma_{i,l}^{p} cB_{z,i+\frac{1}{2}+(l-1)}^{n+1/2}- \sum_{l=1}^{\frac{p}{2}}\psi_{i,l}^{p}cB_{z,i+\frac{1}{2}-l}^{n+\frac{1}{2}}\right]\\
\end{array}&
\end{flalign}

\begin{flalign}
&\begin{array}{lcl}
 cB_{z,i+\frac{1}{2},j+\frac{1}{2},k}^{n+\frac{1}{2}}&=&a^{*}_{i+\frac{1}{2}}cB_{z,i+\frac{1}{2},j+\frac{1}{2}}^{n-\frac{1}{2}}\\
 &-&\beta^{*}_{z,i+\frac{1}{2}}  \left[\sum_{q=1}^{\frac{p}{2}}\Gamma_{i+\frac{1}{2},q}^{p} E_{y,i+q,j+\frac{1}{2}}^{n}-\sum_{q=1}^{\frac{p}{2}}\psi_{i+\frac{1}{2},q}^{p}E_{y,i-(q-1),j+\frac{1}{2}}^{n}\right] \end{array}& \label{eqBzmodi}
\end{flalign}

with: 

\begin{equation}
 a_{i}=\alpha_{y,i}+\frac{\beta_{y,i}\gamma_{x,i}^{*}e^{-j_c\omega \delta t}\left[\sum_{l=1}^{\frac{p}{2}}C_{l}^{p}\left(e^{-j_ck_z(l-1)\delta z}-e^{j_ck_z l \delta z}\right)\right]^2}{\left(e^{j_c\omega \delta t}-\alpha_{y,i+\frac{1}{2}}\right)\left(e^{j_c\omega \delta t}-\alpha^{*}_{x,i}-\beta^{*}_{x,i}e^{j_c(\omega \delta t+k_y\delta y)}\left[\sum_{l=1}^{\frac{p}{2}}C_{l}^{p}\left(e^{-j_ck_y(l-1)\delta y}-e^{j_ck_y l \delta y}\right)\right]^2\right)}
 \end{equation}
and:
\begin{equation}
 a_{i+\frac{1}{2}}^{*}=\alpha^{*}_{z,i+\frac{1}{2}}+\gamma^{*}_{z,i+\frac{1}{2}}\beta_{x,i+\frac{1}{2}}\frac{e^{j_c\omega \delta t}}{e^{j_c\omega \delta t}-\alpha_{x,i+\frac{1}{2}}}\left[\sum_{l=1}^{\frac{p}{2}}C_{l}^{p}\left(e^{-j_ck_y(l-1)\delta y}-e^{j_ck_y l \delta y}\right)\right]^2
 \end{equation}
 
 Equations \ref{eqEymodi} and \ref{eqBzmodi} having the same form as equations (\ref{eqEi}) and (\ref{eqBi}), the method developed with the 1D model is readily applicable for deriving the re-emission coefficients $\xi_l$. When $\theta=0$ and $\phi=0$ we have  $k_y=0$, $k_z=0$  yielding exactly the same equations as in the 1D case previously described. When $\theta\neq 0$ and $\phi\neq 0$, the form of the total electric field $E_y$ and magnetic field $B_z$ is slightly different in the 3D case and is given in the following subsection. 
 
  \subsubsection{Form of electric and magnetic fields in the 3D case}

 It can be shown using the system of equations (\ref{eqExijk}-\ref{eqBzijk}) in vacuum that the relative amplitudes $\chi_{y}=E_{y}/E$ and  $\chi_{z}=B_{z}/B$ where $E$/$B$ are the total electric field/magnetic amplitudes are given by: 
\begin{eqnarray}
\chi_{y}&=&\frac{(1-\epsilon_y^2)^2}{\sqrt{(1-\epsilon_y^2)^4+\epsilon_x^2\epsilon_z^2\epsilon_y^4+(1-\epsilon_y^2)^2\epsilon_s^2\epsilon_y^2}}\\
\chi_{z}&=&\frac{(1-\epsilon_y^2)}{\sqrt{(1-\epsilon_y^2)^2+\epsilon_z^2\epsilon_y^2}}\\
\end{eqnarray}
with $\epsilon_{x},\epsilon_y,\epsilon_z$ satisfying the dispersion relation in vacuum:
\begin{equation}
\epsilon_x^2+\epsilon_y^2+\epsilon_z^2=1 
 \end{equation}
 and given by: 
  \begin{equation}
 \epsilon_{x,y,z}=\frac{\eta_{x,y,z}}{\sin\omega\delta t/2}\sum^{\frac{p}{2}}_{l=1}C^p_{l}\sin \frac{(2l-1)}{2}k_{x,y,z}\delta_{x,y,z}
 \end{equation}

 In the $1D$ case ($k_y=k_z=0$), $\epsilon_y=\epsilon_z=0$ and $\chi_{y}=\chi_{z}=1$. In the 2D case $(k_z=0)$, $\epsilon_z=0$, $\chi_{y}=\epsilon_x$  and $\chi_{z}=1$.

Retaining the same general form for the total electric and magnetic fields $(E_{y,i,j,k}^n,B_{z,i,j,k}^n)$ as in the 1D case:

\begin{eqnarray}
E_{y,i,j,k}^n&=&E_{d,i,j,k}^n+\sum_{l=i_{min}}^{i_{max}}\zeta_{ebynode,i,j,k,l}^{n}+\sum_{l=i_{min}+\frac{1}{2}}^{i_{max}-\frac{1}{2}}\zeta_{ebyinternode,i,j,k,l+\frac{1}{2}}^{n}\label{eqForm:Ey}\\
B_{z,i,j,k}^n&=&B_{d,i,j,k}^n+\sum_{l=i_{min}}^{i_{max}}\zeta_{bbynode,i,j,k,l}^{n}+\sum_{l=i_{min}+\frac{1}{2}}^{i_{max}-\frac{1}{2}}\zeta_{bbyinternode,i,j,k,l+\frac{1}{2}}^{n}\label{eqForm:Bz}\\
\end{eqnarray}

the combination of three components for the electric field $E_y$ (see Fig. \ref{field:form} (a)) is now given by: 
 
 \begin{itemize}

\item the electric field $E_{d,i,j,k}^{n}$ at position $(i,j,k)$ corresponding to the electric field of the driving plane wave: 

\begin{equation}
E_{d,i,j,k}^{n}=\chi_{y} e^{-j_c k_xi\delta x -j_c k_y j\delta y-j_c k_z k\delta z+j_c\omega n \delta t}
\end{equation}

\item the electric field $\zeta_{ebynode,i,j,k,l}^{n}$ at position $(i,j,k)$ emitted by the secondary source plane at position $l$ ($i_{min}<l<i_{max}$):

\begin{equation}
\zeta_{ebynode,i,j,k,l}^{n}=\left \{\begin{array}{cc}
\chi_{y} \xi_l e^{j_c k_x i\delta x-j_c k_y j\delta y-j_c k_z k\delta z+j_c\omega n \delta t}& i\leqslant l\\
\chi_{y} \xi_l e^{-j_c k_x i\delta x-j_c k_y i\delta y-j_c k_z k\delta z+j_c\omega n \delta t}& i\geqslant l\\
\end{array}\right.
\end{equation}

\item the electric field $\zeta_{ebyinternode,i,j,k,l+\frac{1}{2}}^{n}$ at position $(i,j,k)$ emitted by the secondary source plane at internode $l+\frac{1}{2}$ ($i_{min}<l+\frac{1}{2}<i_{max}$):

\begin{equation}
\zeta_{ebyinternode,i,j,k,l+\frac{1}{2}}^{n}=\left \{\begin{array}{cc}
-\chi_{y}  \xi_{l+\frac{1}{2}} e^{j_c k_x i\delta x-j_c k_y i\delta y-j_c k_z k\delta z+j_c\omega n \delta t}& i\leqslant l\\
\chi_{y} \xi_{l+\frac{1}{2}} e^{-j_c k_x i\delta x-j_c k_y i\delta y-j_c k_z k\delta z+j_c\omega n \delta t}& i\geqslant l\\
\end{array}\right.
\end{equation}
\end{itemize}

while the three components for the magnetic field $B_z$ read:

\begin{itemize}

\item the magnetic field $B_{d,i,j,k}^{n}$ at position $(i,j,k)$ corresponding to the electric field of the driving plane wave: 

\begin{equation}
B_{d,i,j,k}^{n}=\chi_z e^{-j_ck_xi\delta x -j_ck_y j\delta y-j_c k_z k\delta z +j_c\omega n \delta t}
\end{equation}

\item the magnetic field $\zeta_{bbyinternode,i,j,k,l+\frac{1}{2}}^{n}$ at position $(i,j,k)$ emitted by the secondary source at internode $l+\frac{1}{2}$ ($i_{min}<l+\frac{1}{2}<i_{max}$):

\begin{equation}
\zeta_{bbyinternode,i,j,k,l+\frac{1}{2}}^{n}=\left \{\begin{array}{cc}
\chi_z\xi_{l+\frac{1}{2}} e^{j_c k_x i\delta x-j_ck_y j\delta y-j_c k_z k\delta z+j_c\omega n \delta t}& i\leqslant l\\
\chi_z\xi_{l+\frac{1}{2}} e^{-j_c k_x i\delta x-j_ck_y j\delta y-j_c k_z k\delta z+j_c\omega n \delta t}& i\geqslant l\\
\end{array}\right.
\end{equation}

\item the magnetic field $\zeta_{bbynode,i,j,k,l}^{n}$ at position $(i,j,k)$ emitted by the secondary source at internode $l$ ($i_{min}<l<i_{max}$):

\begin{equation}
\zeta_{bbynode,i,l}^{n}=\left \{\begin{array}{cc}
-\chi_z \xi_l e^{j_c k_x i\delta x-j_ck_y j\delta y-j_c k_z k\delta z+j_c\omega n \delta t}& i\leqslant l\\
\chi_z \xi_l e^{-j_c k_x i\delta x-j_ck_y j\delta y-j_c k_z k\delta z+j_c\omega n \delta t}& i\geqslant l\\
\end{array}\right.
\end{equation}

\end{itemize}

\section{Numerical validation 1: application of the model to the domain decomposition technique}

We now investigate the effect of domain decomposition on solutions calculated with high-order/pseudo-spectral solvers. 

\subsection{General principle of the domain decomposition technique}

 The principle of the domain decomposition technique  is illustrated on Fig. \ref{DomainDecomp:principle}. For simplicity, we consider first a 1D geometry along $x$. The main domain is split into two semi-infinite subdomains $(D_1)$ for $x<0$ and $(D_2)$ for $x>0$ with $N_{guards}$ guard cells $(G_1)=[0, N_{guards}\delta x]$ for (D1) and $N_{guards}$ guard cells $(G_2)=[-N_{guards}\delta x, 0]$ for $(D_2)$. At each time step $n$, the procedure for exchanging data between the two domains is as follow: 

\begin{figure}[h!]
\centering
\includegraphics[width=0.8\linewidth]{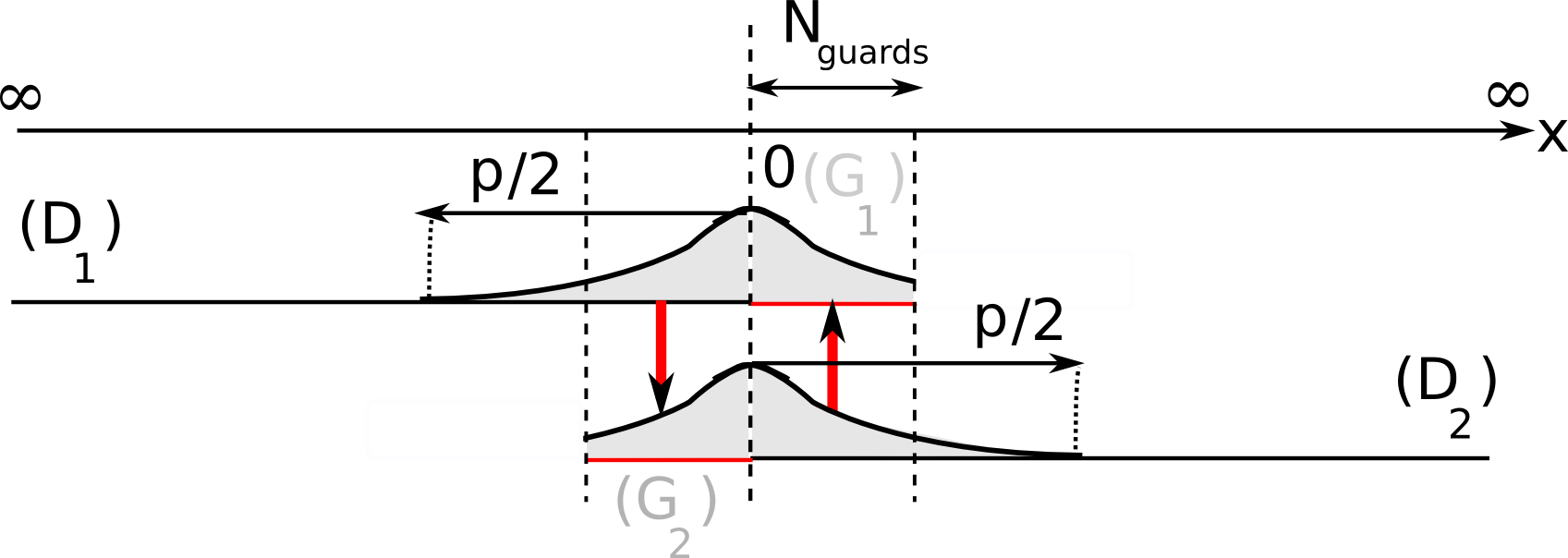}
\caption{Domain decomposition technique. The simulation domain is divided in two semi-infinite subdomains $(D_1)$ and $(D_2)$. The envelope of the stencil of the field solver (order $p$) is represented by a black curve and grey filling between the axis and the curve. At each time step $n$, fields are first updated using Maxwell's equations on each subdomain. Fields in guard regions $(G_1)$ and $(G_2)$ are then replaced by fields at the same positions in $(D_2)$ and $(D_1)$ respectively (red arrows).  }
\label{DomainDecomp:principle}
\end{figure}

\begin{enumerate}[(i)]
\item the electromagnetic field component is updated using Maxwell's equations independently on each subdomain $(D_1)$ and $(D_2)$ (Fields in $(G_1)$ and $(G_2)$ are not updated), 
\item nodes from $(D_2)$ at same positions as nodes in $(G_1)$ are copied to $(G_1)$, 
\item nodes from $(D_1)$ at same positions as nodes in $(G_2)$ are copied to $(G_2)$. 
\end{enumerate}

This procedure is done for every field component. Using finite-difference leapfrog time integration,  $\textbf{B}^{n-1/2}$ is first advanced to $\textbf{B}^{n+1/2}$ and copied to adjacent guard regions. $\textbf{E}^{n}$ is then advanced to $\textbf{E}^{n+1}$ using $\textbf{B}^{n+1/2}$ and copied to adjacent guard regions. 
 If $N_{guards}\geqslant p/2$, domain decomposition will not affect the precision of the scheme and the result will be equal to the one on a single grid without domain decomposition. 

Because exchanging large volumes of data can be prohibitively expensive in terms of computational ressources on massively parallel supercomputers, it has been proposed to limit the volume of data exchanges between subdomains, and thus the number of guard cells for high-order stencil, or its infinite order pseudo-spectral limit \cite{Vay2013}. However, errors are introduced when $N_{guards}<p/2$. In this case,  $p/2-N_{guards}$ nodes are affected in $(D_1)$ and $(D_2)$ by truncation of the stencil. This eventually leads to spurious signals in each subdomain and potential numerical errors. In the following we use our model to compute the total error generated by field truncations and exchange at the boundaries and compare theoretical results to simulations for plane waves at normal incidence or at an angle from the domains' interface. 

\subsection{Domain decomposition strategies for staggered field solvers}

\begin{figure}[h!]
\centering
\includegraphics[width=0.8\linewidth]{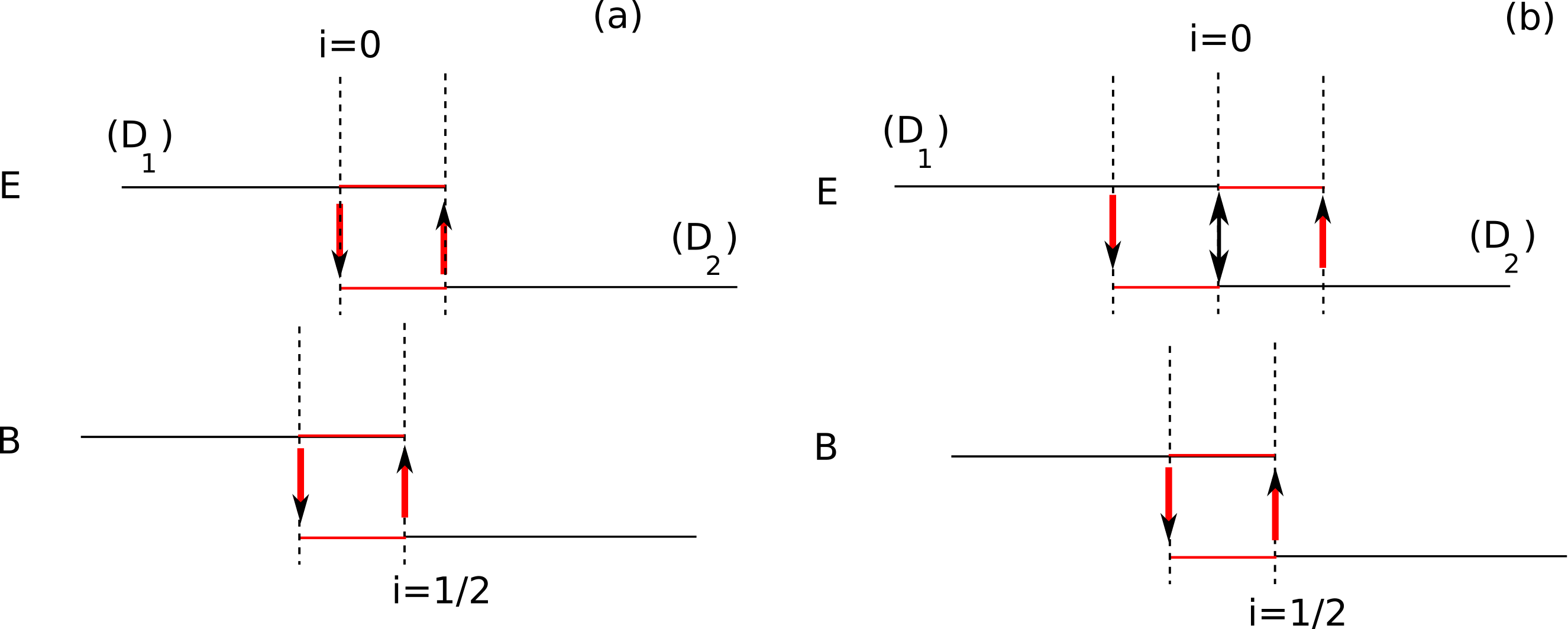}
\caption{Domain decomposition strategies. (a) First domain decomposition strategy (Strategy $1$). $E$ and $B$ fields are exchanged in a staggered way. (b) Second domain decomposition strategy (Strategy $2$) where $E$ and $B$ are exchanged in a centered way. The value of the electric field at the central node $i=0$ is averaged by taking $E_{i=0}^{n+1}=(E_{i=0,D_1}^{n+1}+E_{i=0,D_2}^{n+1})/2$, where $E_{i=0,D_1}^{n+1}$ (resp. $E_{i=0,D_2}^{n+1}$) is the E-field value calculated on $D_1$ (resp. $D_2$) at $i=0$. }
\label{DomainDecomp:strat}
\end{figure}

In the numerical validations of our model, we will compare two domain decomposition techniques for staggered field solvers. The first one (staggered method) is sketched on Fig. \ref{DomainDecomp:strat} (a) for a 1D geometry. $E$ and $B$ fields are exchanged in a staggered way. 

The second technique (centered method) consists of centering the exchange of $E$ and $B$ fields by overlapping nodes from $(D_1)$ and $(D_2)$ (see Fig. \ref{DomainDecomp:strat} (b)). In that case, the value of the electric field at the central node $i=0$ is averaged by taking $E_{i=0}^{n+1}=(E_{i=0,D_1}^{n+1}+E_{i=0,D_2}^{n+1})/2$, where $E_{i=0,D_1}^{n+1}$ (resp. $E_{i=0,D_2}^{n+1}$) is the E-field value calculated on $D_1$ (resp. $D_2$) at $i=0$. In the following we will refer to the staggered method as "Strategy -$1$" and the centered method as "Strategy-$2$".

\subsection{Practical implementation of the domain decomposition technique in our model}

For "Strategy-$1$", the domain decomposition technique is accounted for using the following coefficients in Maxwell's equations (\ref{eqEi}, \ref{eqBi}) for the 1D model and (\ref{eqEymodi}, \ref{eqBzmodi}) for the 3D model: 

\begin{equation}
\Gamma_{i,l}^{p}=
\left \{\begin{array}{ccc}
&C_{l,p} & i>0 \text{ .or. } p/2<N_{guards}\\
&C_{l,p} & i<0\text{ .and. }i+l\leqslant N_{guards}\\
&0.& i<0\text{ .and. }i+l> N_{guards}\\
&C_{l,p} & i=0\text{ .and. }i+l\leqslant N_{guards}\\
&0.& i=0\text{ .and. }i+l> N_{guards}\\
\end{array}\right.
\end{equation}

\begin{equation}
\psi_{i,l}^{p}=
\left \{\begin{array}{ccc}
&C_{l,p} & i\leqslant 0 \text{ .or. } p/2<N_{guards}\\
&C_{l,p} & i>0\text{ .and. }i-l\geqslant -N_{guards}\\
&0.& i>0\text{ .and. }i-l< N_{guards}\\
\end{array}\right.
\end{equation}

For "Strategy-$2$", the domain decomposition technique is accounted for using the following coefficients:

\begin{equation}
\Gamma_{i,l}^{p}=
\left \{\begin{array}{ccc}
&C_{l,p} & i>0 \text{ .or. } p/2<N_{guards}\\
&C_{l,p} & i<0\text{ .and. }i+l\leqslant N_{guards}\\
&0.& i<0\text{ .and. }i+l> N_{guards}\\
&C_{l,p}/2 & i=0\text{ .and. }i+l\leqslant N_{guards}\\
&0.& i=0\text{ .and. }i+l> N_{guards}\\
\end{array}\right.
\end{equation}

\begin{equation}
\psi_{i,l}^{p}=
\left \{\begin{array}{ccc}
&C_{l,p} & i<0 \text{ .or. } p/2<N_{guards}\\
&C_{l,p} & i>0\text{ .and. }i-l\geqslant -N_{guards}\\
&0.& i>0\text{ .and. }i-l< N_{guards}\\
&C_{l,p}/2 & i=0\text{ .and. }i-l\geqslant -N_{guards}\\
&0.& i=0\text{ .and. }i-l< N_{guards}\\
\end{array}\right.
\end{equation}

Besides, in both cases we have: 

\begin{equation}
\alpha_{x,i}=\alpha^{*}_{x,i}=\alpha_{y,i}=\alpha_{z,i}=\alpha^{*}_{z,i}=1
\end{equation}
and 

\begin{eqnarray}
\beta^{*}_{z,i}&=&\beta_{z,i}=\gamma^{*}_{y,i}=\gamma_{y,i}=\eta_{x}\\
\beta_{x,i}&=&\beta^{*}_{x,i}=\gamma^{*}_{z,i}=\eta_{y}\\
\gamma^{*}_{x,i}&=&\beta_{y,i}=\eta_{z}\\
\end{eqnarray}

\subsection{Theory-simulation comparisons}

In this section, we validate our model against simulations in the case of a plane monochromatic wave of frequency $\omega$ impinging at oblique incidence $(\theta\neq 0^o,\phi=0^o)$ on the subdomain boundary. 

\subsubsection{Test procedure and numerical error measurements}

The simulation domain is divided into two subdomains of equal sizes with $N_{guards}$ cells at their boundaries. At each time step, fields are computed independently on each subdomain and guard cells are exchanged between subdomains.  At $t=0$, a Harris-like waveform $H(t,y)$ is launched at the left boundary of the simulation domain: 

\begin{equation}
H(t,y,\lambda)=h(t)\sin\left(\omega t-k y\sin\theta\right)
\end{equation}

where $\theta$ is the angle of incidence of the waveform, $k=2\pi/\lambda$ the wavenumber and $\omega$ the angular frequency obtained by the numerical dispersion of the staggered scheme. Amplitude $h(t)$ is the Harris function given by: 

\begin{equation}
h(t)=\left \{\begin{array}{cc}
\frac{1}{32}\left[10-15\cos\frac{2\pi t}{T}+6\cos\frac{4\pi t}{T}-\cos\frac{6\pi t}{T}\right]&0<t\leqslant T\\
0 &t>T\\
\end{array}\right.
\end{equation}

This wave-form has a quasi-monochromatic spectrum, enabling validation of the model at specific wavelength $\lambda$. The modulus of the numerical re-emission coefficient $|\zeta_{sim}|$ is obtained by taking the ratio of the reflected field energy $R$ over the incident wave energy $I$:

\begin{equation} 
|\zeta_{sim}|=\sqrt{\frac{R}{I}}=\sqrt{\frac{\sum_{Left}(E^2+c^2 B^2)}{I}}
\end{equation}

where the sum $\sum$ in the above equation is taken on the left subdomain and at the end of the simulation. In order to obtain also the phase properties of the secondary sources in the simulations, we compute $|1-\zeta_{sim}|$, which equates the ratio of transmitted energy $T$ over the incident wave energy $I$: 
\begin{equation} 
|1-\zeta_{sim}|=\sqrt{\frac{T}{I}}=\sqrt{\frac{\sum_{Right}(E^2+c^2 B^2)}{I}}
\end{equation}
where the sum $\sum$ in the above equation is taken on the right subdomain and at the end of the simulation. The phases $Arg(\zeta_{sim})$ and $Arg(1-\zeta_{sim})$ of the reflected and transmitted waves in these simulation are finally given by: 

\begin{eqnarray}
Arg(\zeta_{sim})&=&\arccos\left[\frac{1+|\zeta_{sim}|^2-|1-\zeta_{sim}|^2}{2|\zeta_{sim}|}\right]\\
Arg(1-\zeta_{sim})&=&\arccos\left[\frac{1+|1-\zeta_{sim}|^2-|\zeta_{sim}|^2}{2|1-\zeta_{sim}|}\right]\\
\end{eqnarray}

In the following paragraphs, we compare, for different numerical parameters, the coefficients $|\zeta_{sim}|$, $|1-\zeta_{sim}|$, $Arg(\zeta_{sim})$ and $Arg(1-\zeta_{sim})$ to our theoretical estimates $|\zeta_{th}|$, $|1-\zeta_{th}|$, $Arg(\zeta_{th})$ and $Arg(1-\zeta_{th})$  provided by the model described in section \ref{sec:model}. 

\subsubsection{Influence of order $p$ and number of guard cells $N_{guards}$}

\begin{figure}[h!]
\centering
\includegraphics[width=0.6\linewidth]{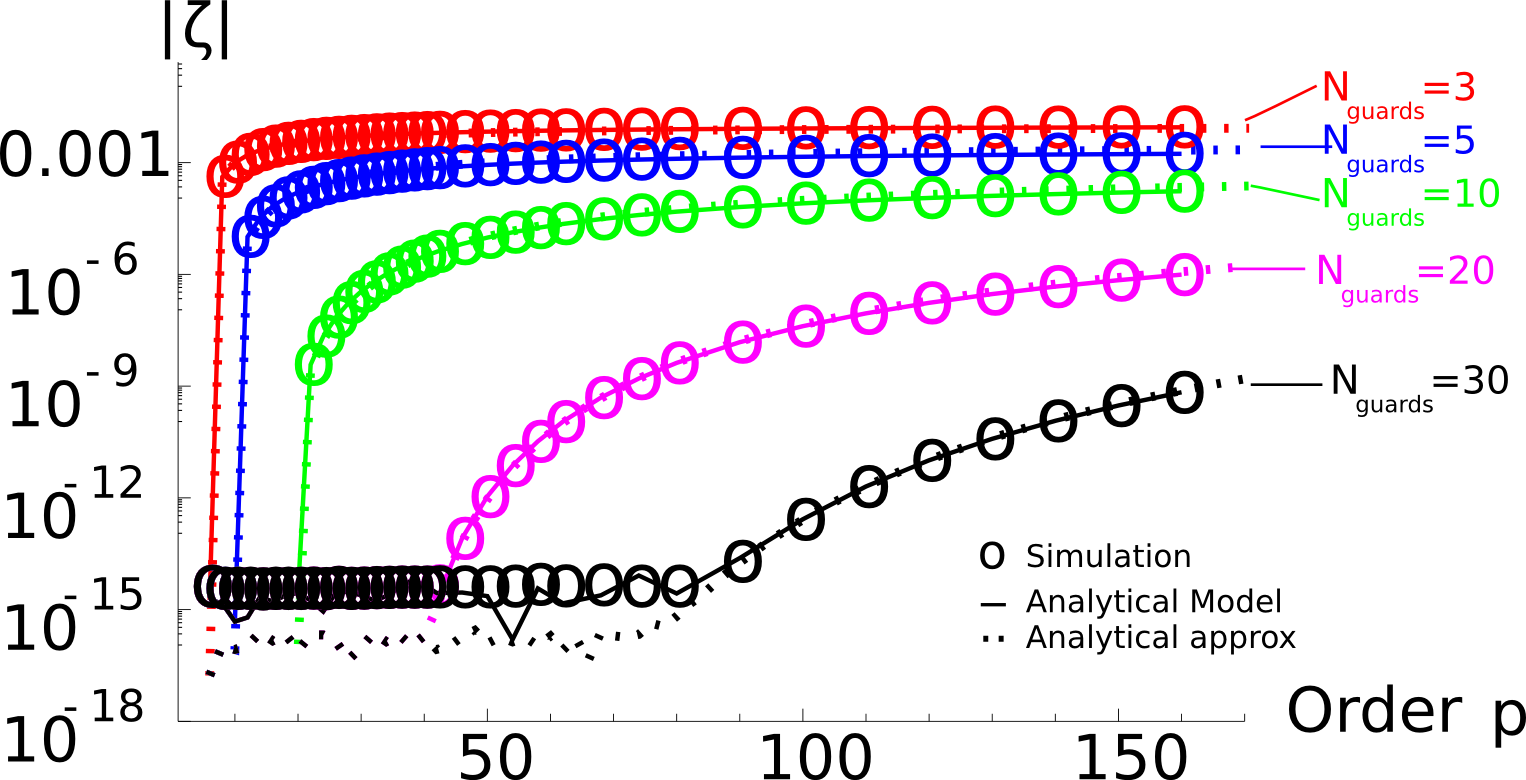}
\caption{ Variation of the truncation error $\zeta$ (total re-emission coefficient) with order $p$ for a different number of guard cells $N_{guards}=3,5,10,20,30$ corresponding to red, blue, green, magenta and black curves. $\lambda/\delta x=10$,  and $(\theta=0^o,\phi=0^o)$ are fixed. "Strategy-2" was used as domain decomposition technique. For each curve, circles represent the re-emission coefficient $\zeta$ calculated in the simulations, solid lines represent $\zeta$ computed with our model and dotted lines represent $|\zeta|$ deduced from the analytical approximate formula derived in \ref{anapproxp}.}
\label{fig:Modelvsorder}
\end{figure}

Figure \ref{fig:Modelvsorder} (a) represents the variation of the modulus of the total re-emission coefficient $|\zeta|$ with order $p$ of the Maxwell solver for different number of guard cells $N_{guards}$ (see colored curved). The agreement between the model (solid lines) and simulations (circles) is excellent. When $N_{guards}\geqslant p/2$, there is no stencil truncation and $|\zeta|$ is zero at double machine precision $\chi_{double}\approx 10^{-15}$. However, when  $N_{guards}<p/2$ the stencil is truncated at subdomain boundaries and a spurious signal of amplitude $|\zeta|$ is created. 

For a given number of guard cells $N_{guards}$, the stencil truncations increase when order $p$ increases and the amplitude $\zeta$ also becomes larger. Figure \ref{fig:Modelvsorder} shows that $\zeta$ seems to converge to a a constant value when $p\gg 1$ allowing the estimation of $\zeta_{\infty}$ when $p\rightarrow\infty$ i.e when the FDTD Maxwell solver turns into a pseudo-spectral solver. 

On the contrary, when $N_{guards}$ increases for a given oder p, stencil truncations are reduced and  $|\zeta|$ decreases, as expected. 

\begin{figure}[h!]
\centering
\includegraphics[width=0.99\linewidth]{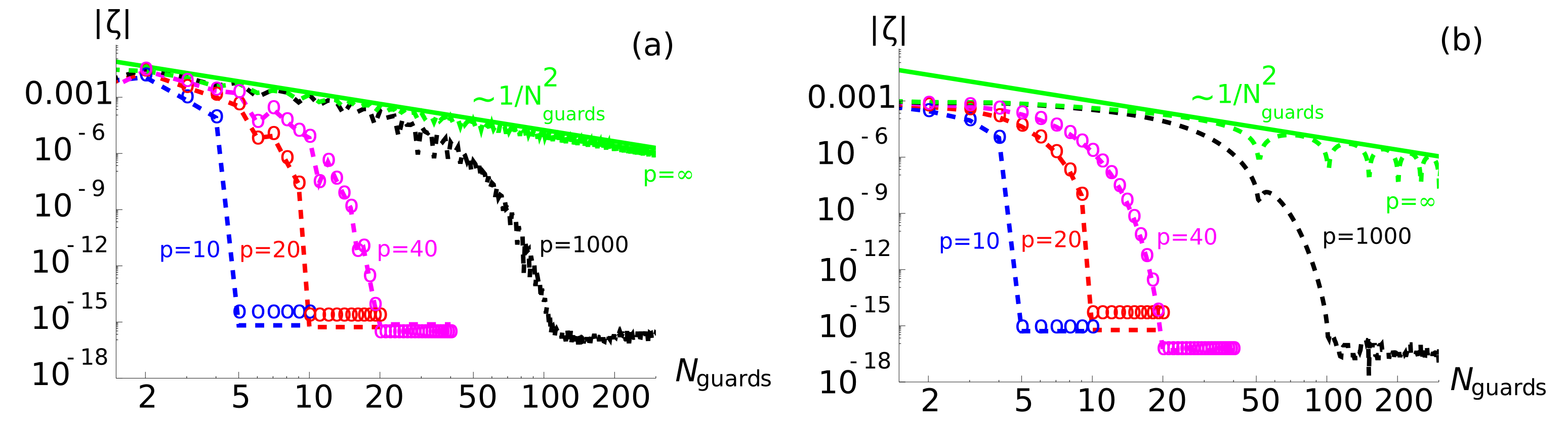}
\caption{ Variation of the truncation error $\zeta$ (total re-emission coefficient) with the number of guard cells for different orders $p$ and wavelengths $\lambda/\delta x$. "Strategy-2" was used as domain decomposition technique. $\eta_{x}=0.4$ and $(\theta=0^o,\phi=0^o)$ are fixed. For each curve, dashed lines represent the re-emission coefficient $|\zeta|$ deduced from the analytical approximate formula (cf. \ref{anapproxp}). Red, Blue and Magenta circles represent $|\zeta|$ computed with the full analytical model. The green solid line represents the function $0.18/N_{guards}^2$. (a) Variation of $|\zeta|$ with $N_{guards}$ for $\lambda/\delta x=5$  (b) Variation of $|\zeta|$ with $N_{guards}$ for $\lambda/\delta x=100$.  }
\label{fig:ApproximatevsNguard}
\end{figure}

Note that it is possible to get a practical analytical approximate formula of the error amplitude $|\zeta|$ (cf. \ref{anapproxp}). This estimate is represented by dotted lines on Fig. \ref{fig:Modelvsorder} and dashed lines on Fig. \ref{fig:ApproximatevsNguard}. It very accurately reproduces the evolution of $|\zeta|$ with $p$, $N_{guards}$ and $\delta x/\lambda$. 

The limit $|\zeta|_{\infty}$ of this analytical approximation when $p\rightarrow \infty$ is derived in \ref{anapproxinfinite}. The variation of the error $|\zeta|_{\infty}$ with the number of guard cells $N_{guards}$ is represented in Fig. \ref{fig:ApproximatevsNguard} (a) and (b) (green dashed line). The maximum error (green line) varies as $1/N_{guards}^2$ when $N_{guards}\gg 1$ which could have been guessed by simply noticing that the amplitude of the stencil coefficients $|C_{l}^{\infty}|$ vary as $1/(2l-1)^2$. This means that taking $10$ times more guard cells yields a maximum error divided by $100$. Even if the absolute error is relatively low, having zero error at machine precision ($\approx10^{-15}$ for double precision) would mandate too many guard cells. Fig. \ref{fig:ApproximatevsNguard} shows however that for orders as high as $p=1000$, the number of guard cells needed to have zero error at machine precision remains very low ($N_{guards}\approx 100$ for $p=1000$). As order $p=1000$ already yields spectral resolution at machine precision on a large band of frequencies, it would thus be more interesting to use very high finite order solvers instead of infinite order solvers with domain decomposition. These results will be presented in greater details in further work. 


\subsubsection{Influence of wavelength $\lambda$ of the driving wave}

\begin{figure}[h!]
\centering
\includegraphics[width=0.99\linewidth]{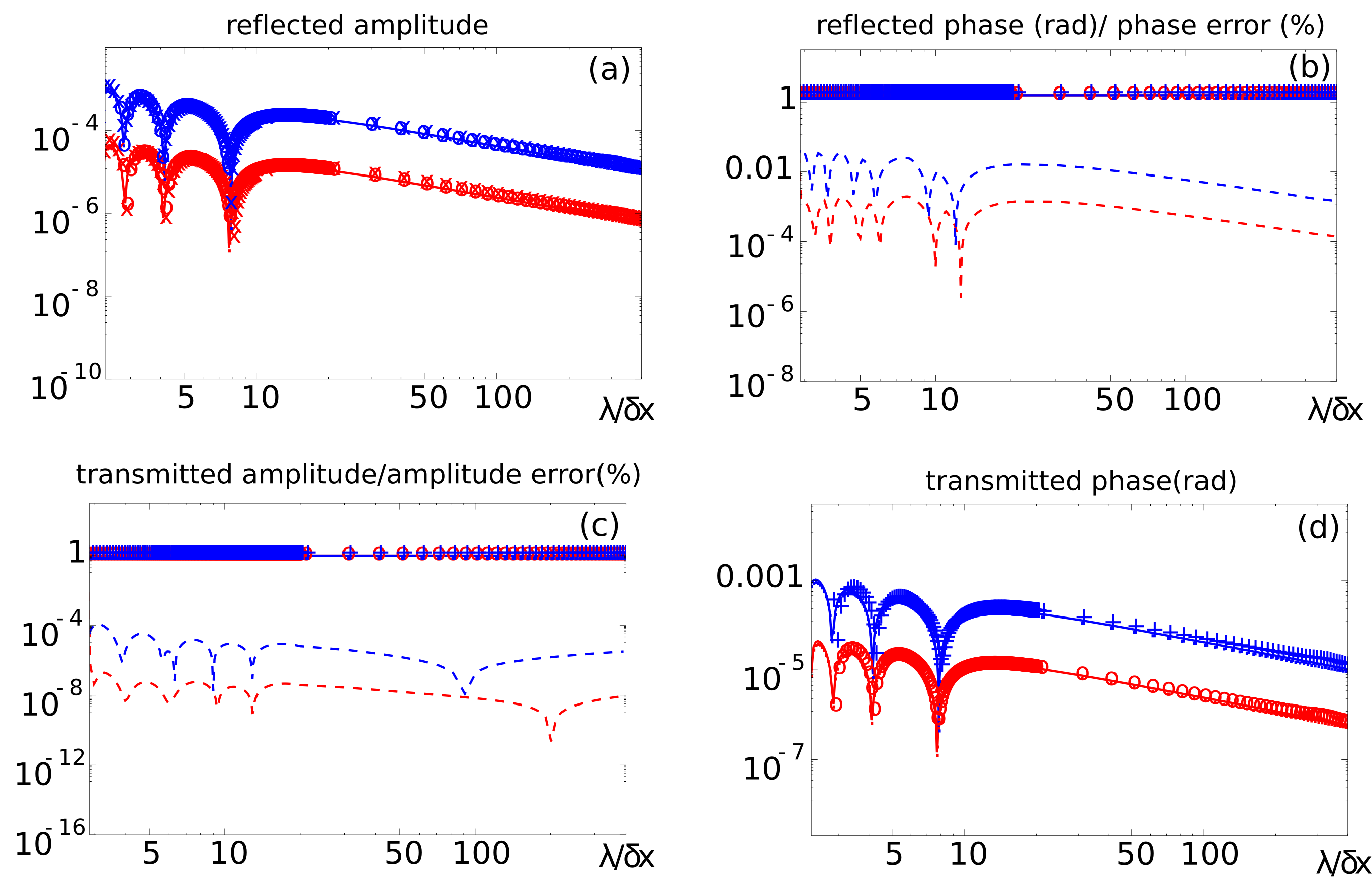}
\caption{Variation of the truncation signal amplitudes and phases with $\delta x/\lambda$. Parameters $N_g=5$, $\eta_x=0.4$ and $\theta=\phi=0^o$ are fixed. "Strategy-2" was used as domain decomposition technique. The blue lines correspond to the case $p=20$ and the red lines to $p=12$. In each plot, solid lines correspond to results of the model. "+" and "o" markers correspond to simulation results. "$\times$" markers correspond to $|\zeta|$ deduced from the analytical approximate formula (cf. \ref{anapproxp}) (a) Variation of the amplitude $|\zeta|$ of the reflected signal as a function of the ratio $\lambda/\delta x$. (b) Variation of the phase $Arg(\zeta)$ of the reflected signal as a function of the ratio  $\lambda/\delta x$. The dashed lines correspond to relative errors in $\%$ between the model and simulations for $p=20$ (blue lines) and $p=8$ (red line). (c) Variation of the amplitude $|1-\zeta|$ of the transmitted signal as a function of $\lambda/\delta x$. The dashed lines correspond to relative errors in $\%$ between the model and simulations for $p=20$ (blue lines) and $p=8$ (red line). (d) Variation of the phase $Arg(1-\zeta)$ of the transmitted signal as a function of $\lambda/\delta x$. }
\label{fig:Modelvslambda}
\end{figure}

The model is now compared to simulations when $\lambda/\delta x$ is varied for fixed parameters $N_{guards}$, $\eta_x=0.4$ and $\theta=\phi=0^o$. Results of the model and simulations are represented in Fig. \ref{fig:Modelvslambda}. Our analytical model again perfectly reproduces simulation results. The analytical approximate ($x$ markers in Fig. \ref{fig:Modelvslambda} (a)) derived in \ref{anapproxp} again yields very accurate estimates of the error magnitude. 

 The variation of the modulus $|\zeta|$ and phase $Arg(\zeta)$ of the total re-emission coefficient $\zeta$ are represented on panels (a) and (b). As expected $\zeta$ decreases for longer wavelength. This can be qualitatively understood by considering the fact that for long wavelengths $\lambda$, the stencil is truncated on a region that becomes spatially very small compared to $\lambda$. For any wavelength, the emitted field in the $x<0$ direction is dephased by a constant $\pi/2$ from the driving field, which is actually equivalent to a reflection of the incident field on the subdomain boundary with a reflection coefficient $\zeta$. 

The modulus $|1-\zeta|$ and phase $Arg(1-\zeta)$ of the transmitted wave in  the $x>0$ direction are represented on panels (c) and (d). Panel (c) shows that the amplitude $|1-\zeta|$ of this wave is approximately equals to the amplitude of the driving wave on the whole frequency domain and with a very low dephasing. Consequently, there is low effect of stencil truncations on the wave passing through subdomain boundaries. 

However, our model shows that energy created in the simulation box $\Delta E$ after boundary crossing is given by: 

\begin{eqnarray}
\Delta E/E_{inc}&=&(|1-\zeta|+|\zeta|)-1\\
&=&|1-|\zeta|e^{j_cArg(\zeta)}|+|\zeta|-1\\
&=&|1-j_c|\zeta||+|\zeta|-1\\
&=&\sqrt{1+|\zeta|^2}+|\zeta|-1\\
&=&1+\frac{|\zeta|^2}{2}+|\zeta|-1+o(|\zeta|^2)\\
&=&|\zeta|+o(|\zeta|)
\label{deltae}
\end{eqnarray}

where we used $|\zeta|\ll 1$ (cf. Fig. \ref{fig:Modelvslambda} (a)), $Arg(\zeta)\approx \pi/2$ (cf. Fig. \ref{fig:Modelvslambda} (b)) and $E_{inc}$  the initial energy of the driving wave. 

The total energy after boundary crossing is thus now greater than the initial energy of the driving wave and equals to $|\zeta|$. This means that stencil truncations created non-physical energy in the simulation box of magnitude $|\zeta|$ (cf. Fig. \ref{fig:Modelvslambda} (a)). If the driving pulse crosses several boundaries during the simulation, the total energy will thus increase through time with a growth rate that is a function of the number of crossed subdomains $N_{sub}$ and $|\zeta|$. Moreover, if the number of crossed subdomains is high so that $N_{sub}Arg(1-\zeta)\approx 1$ dephasing effects could also start to alter the phase of the transmitted wave. These effects are under study and further analyses will be presented in greater details elsewhere.

\subsubsection{Influence of the angle of incidence $\theta$}

\begin{figure}[h!]
\centering
\includegraphics[width=0.5\linewidth]{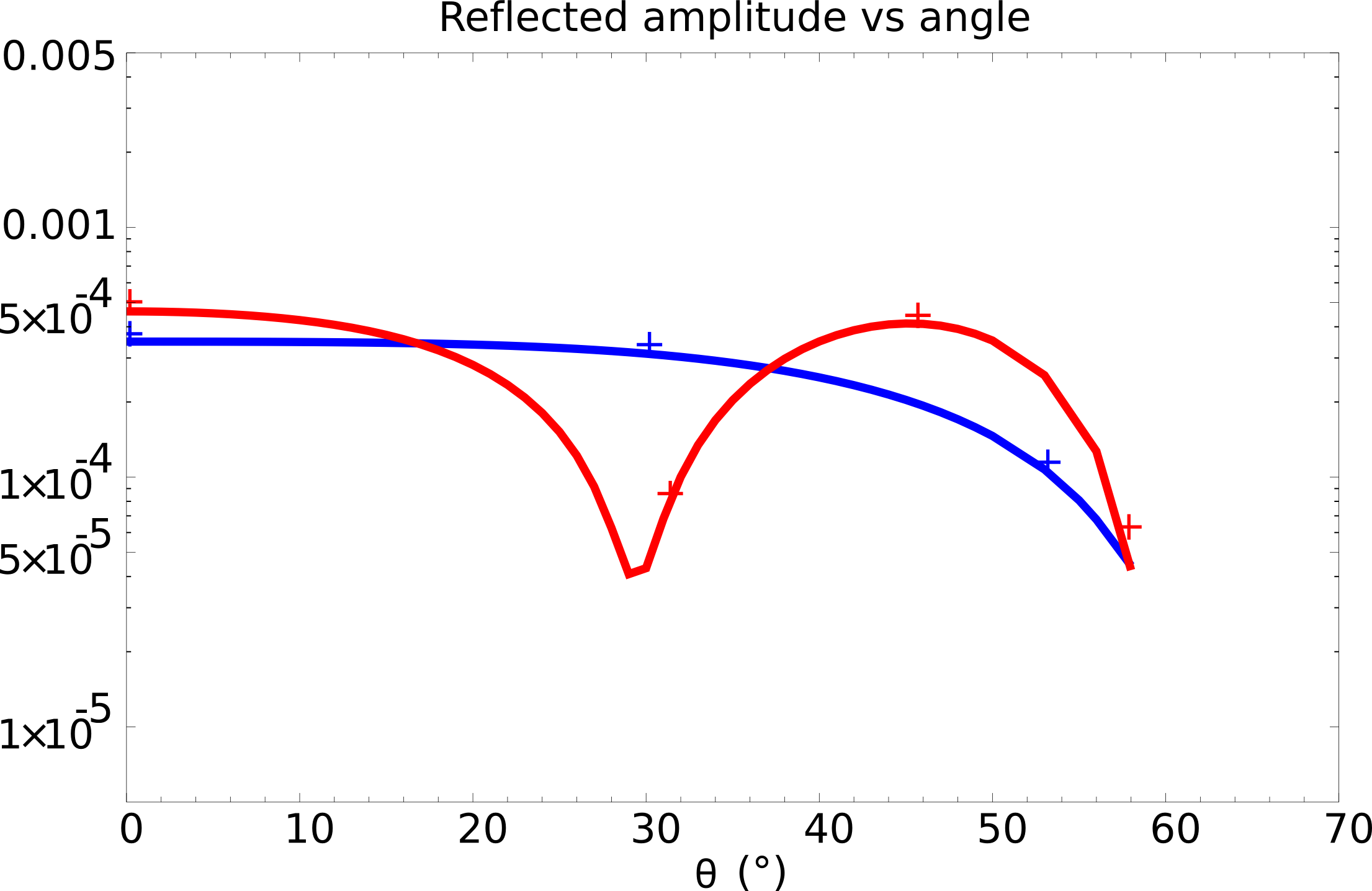}
\caption{Variation of the truncation signal amplitude (re-emission coefficient) $|\zeta|$  with the angle of incidence $\theta$ for the two  different $\lambda/\delta x=3.4$ (red curves) and $\lambda/\delta x=10$ (blue curves). Parameters $N_g=5$, $\eta_x=0.4$ and  order $p=20$.  "Strategy-1" was used as domain decomposition technique. Solid lines represent results of the model and points results from simulations.}
\label{fig:zetavstheta}
\end{figure}

Fig. \ref{fig:zetavstheta} shows the variation of the modulus of the spurious reflected signal (relative to the incident wave modulus) $|\zeta|$ with the angle of incidence $\theta$ for two different wavelengths  $\lambda/\delta x=3.4$ (red curves) and $\lambda/\delta x=10$ (blue curves). In all cases, it appears that variations of $|\zeta|$ with $\theta$ are small for low angles. 

Notice that in our case, there is no stencil variations along directions y (index j) and z (index k). This implies that the spatial phase shift induced by the domain decomposition is independent of $j$ and $k$. As a consequence, there is no spread in propagation direction of the reflected/transmitted waves, which is in accordance with the initial assumptions of the 3D model. 

\subsubsection{Influence of the domain decomposition strategy}

\begin{figure}[h!]
\centering
\includegraphics[width=0.42\linewidth]{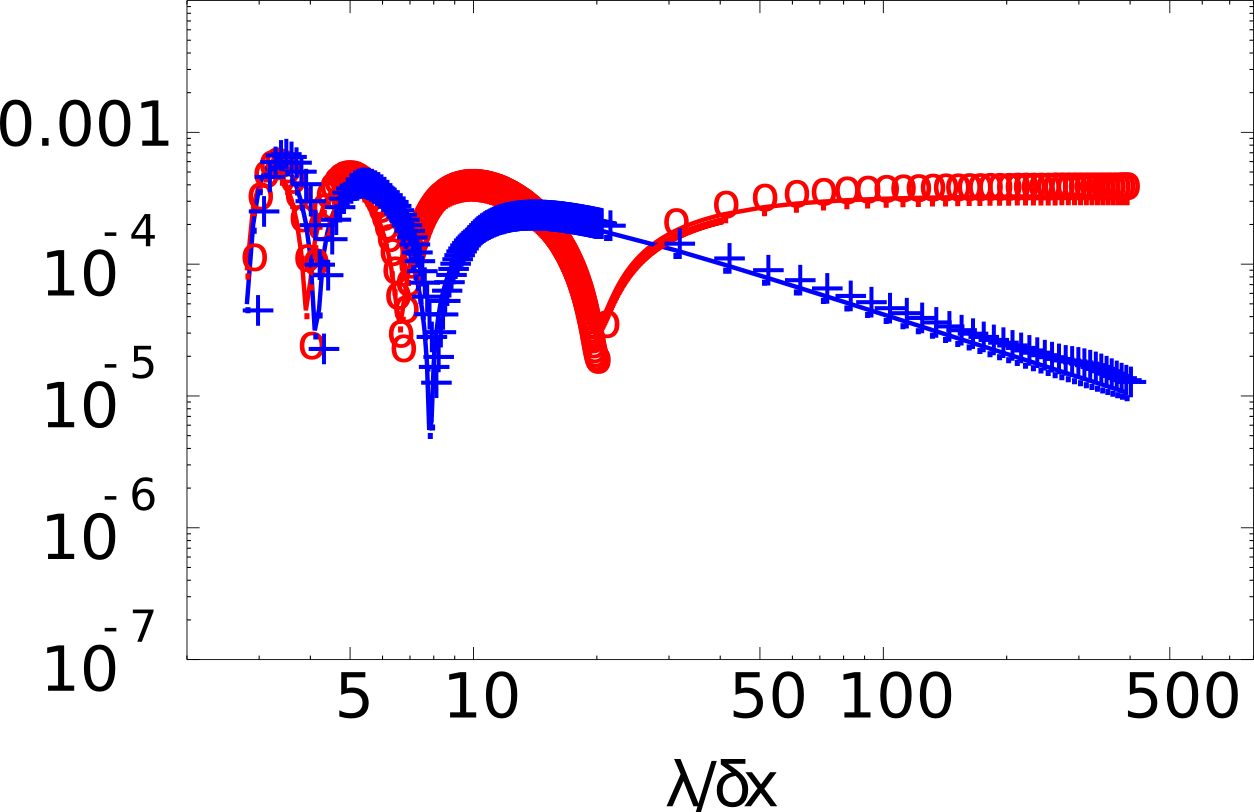}
\caption{Variation of the truncation signal amplitude $|\zeta|$  with $\delta x/\lambda$ for the two different domain decomposition strategies. Parameters $N_g=5$, $\eta_x=0.4$, order $p=20$ and $\theta=\phi=0^o$ are fixed.  Red curves correspond to "Strategy-1" and blue curves to "Strategy-2". Solid lines represent results of the model and points results from simulations.}
\label{fig:ModelvsStrategy}
\end{figure}

Fig. \ref{fig:ModelvsStrategy} illustrates the effect of the domain decomposition method on the re-emission coefficient $|\zeta|$. In the case of "Strategy-1" (red curves), $|\zeta|$ remains fairly uniform on the whole spectral range. On the contrary, in the case of "Strategy-2" (blue curves)  $|\zeta|$ decreases with $\lambda/\delta x$ and is significantly lower than with "Strategy-1" at large $\lambda/\delta x$ and seems to vanish as $\lambda \rightarrow\infty $. This indicates that Strategy-1 method induces a systematic error that is not present with Strategy-2 and that the latter is thus preferable.

\section{Numerical validation 2: application of the model to Perfectly Matched Layers}

In this section we illustrate the versatility of the new analytical model through its application to the evaluation of the reemission coefficient $|\zeta|$ of a PML. Former studies noticed a discrepancy between theory and simulations using a 'one-source' model at high orders $p$ \cite{Lee2015}. In the following, simulations and results of this 'one-source' model are compared to our "multi-sources" model.

\subsection{Practical implementation of a PML in our model}

Numerical parameters for the PML are the same as in \cite{Lee2015}:
\begin{itemize}
\item electric/magnetic conductivities in the PML of $\sigma(i)=\sigma_{max}(i\delta x/\Delta)^2$ with $\sigma_{max}=4/\delta x$ and $\Delta=5\delta x$, 
\item the PML is made of $N_{pml}=20$ cells. 
\end{itemize}

As shown in Appendix B, a PML (using Bérenger's original split formulation in this example) is described in our model by using the following coefficients (see Appendix B for detailed calculations): 

\begin{equation}
\psi_{i,l}^{p}=\Gamma_{i,l}^{p}=C_{l,p}, \forall i
\end{equation}

with: 

\begin{eqnarray}
\alpha_{x,i+\frac{1}{2}}&=&\alpha_{z,i}=\alpha^*_{x,i}=1\\
\alpha_{y,i}&=&\alpha(i)\\
\alpha^*_{z,i+\frac{1}{2}}&=&1\\
\end{eqnarray}

as well as: 

\begin{eqnarray}
\gamma_{y,i}&=&\eta_{x}\\
\beta_{x,i+\frac{1}{2}}&=&\beta_{x,i}^{*}=\beta_{z,i}=\gamma^{*}_{z,i+\frac{1}{2}}=\eta_{y}\\
\gamma^{*}_{x,i}&=&\eta_{z}\\
\beta_{y,i}&=&\beta(i)\\
\beta^*_{z,i+\frac{1}{2}}&=&\beta(i+\frac{1}{2})\gamma(i+\frac{1}{2})\\
\end{eqnarray}

and : 

\begin{eqnarray}
\alpha(i)&=&\frac{2-\sigma(i)\delta t}{2+\sigma(i)\delta t}\\
\beta(i)&=&\frac{2c^2}{2+\sigma(i)\delta t}\eta_x\\
\gamma(i)&=&\frac{e^{j_c\omega\delta t/2}-e^{-j_c\omega\delta t/2}}{e^{j_c\omega\delta t/2}-\alpha(i)e^{-j_c\omega\delta t/2}}
\end{eqnarray}

In the following, our "multi-sources" model is used to compute the total re-emission coefficient $\zeta$ from a PML and is compared to theoretical and simulation results of \cite{Lee2015} that uses the "single-source" model. 

\subsection{Influence of wavelength $\lambda$ on $|\zeta|$}

\begin{center}
\begin{figure}[!h]
\centering
\includegraphics[width=\linewidth]{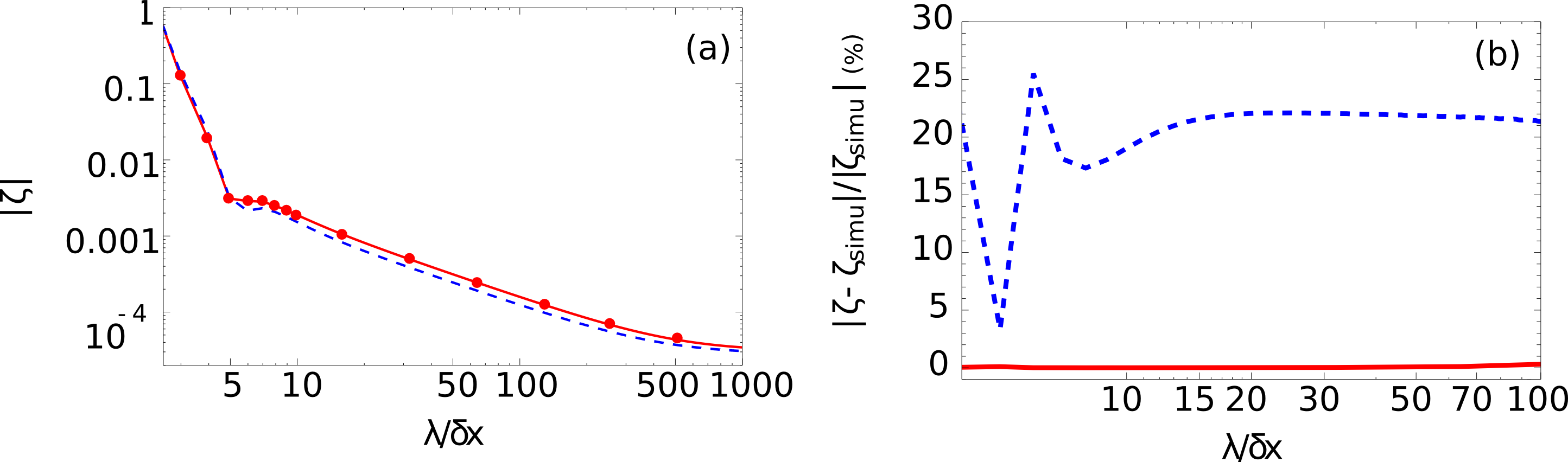}
\caption{ Total re-emission coefficient of a PML $\zeta$. (a) Comparison of the total re-emission coefficient $\zeta$ computed with our analytical model (red line) and with 1D numerical simulations (red dots). Fixed parameters are the courant parameter $\eta=\delta t/\delta x=0.4$, angle of incidence $(\theta=0,\phi=0)$ and the order of the field solver $p=8$. The dashed blue line represents the reflection coefficient computed with the "1-source" model developed in \cite{Lee2015} (b) The ref line correspond to the relative error $|\zeta-\zeta_{simu}|/|\zeta_{simu}|$ in $\%$ between  $\zeta$ calculated by our "p-source" model and $\zeta_{simu}$  calculated by numerical simulations. The red line correspond to  the relative error $|\zeta-\zeta_{simu}|/\zeta_{simu}$ in $\%$ between  $\zeta$ calculated by the "1-source" model as developed in  \cite{Lee2015} and $\zeta_{simu}$  calculated by numerical simulations.}
\label{examplePML}
\end{figure}
\end{center}

Fig. \ref{examplePML}, represents the variation of $|\zeta|$ with the wavelength  $\lambda/\delta x$. Panels (a) and (b) of Fig. \ref{examplePML} show that the new "multi-sources" model agrees perfectly with simulation results, validating its predictive value. Furthermore, it explains and solves the origin of the discrepancy observed between the one-source model and simulations at high orders. Indeed, Panel (b) shows a disagreement\footnote{Notice that the discrepancy between the "1-source" model and simulations seem lower than $20\%$ on panel (a) because data are plotted in log-scale. } of about $20\%$ between the "1-source" model and the simulations results at order $p=8$ (blue dashed line). As explained before, this mismatch between theory and simulation comes from the fact that previous models only considered one secondary source at high orders. Our model however solves this discrepancy (red line on panel (b)) by including all the secondary sources that also radiate and couple each other to contribute to the total re-emission coefficient $|\zeta|$.

\subsection{Influence of angle of incidence $\theta$ of the driving wave on the PML}

\begin{center}
\begin{figure}[!h]
\centering
\includegraphics[width=\linewidth]{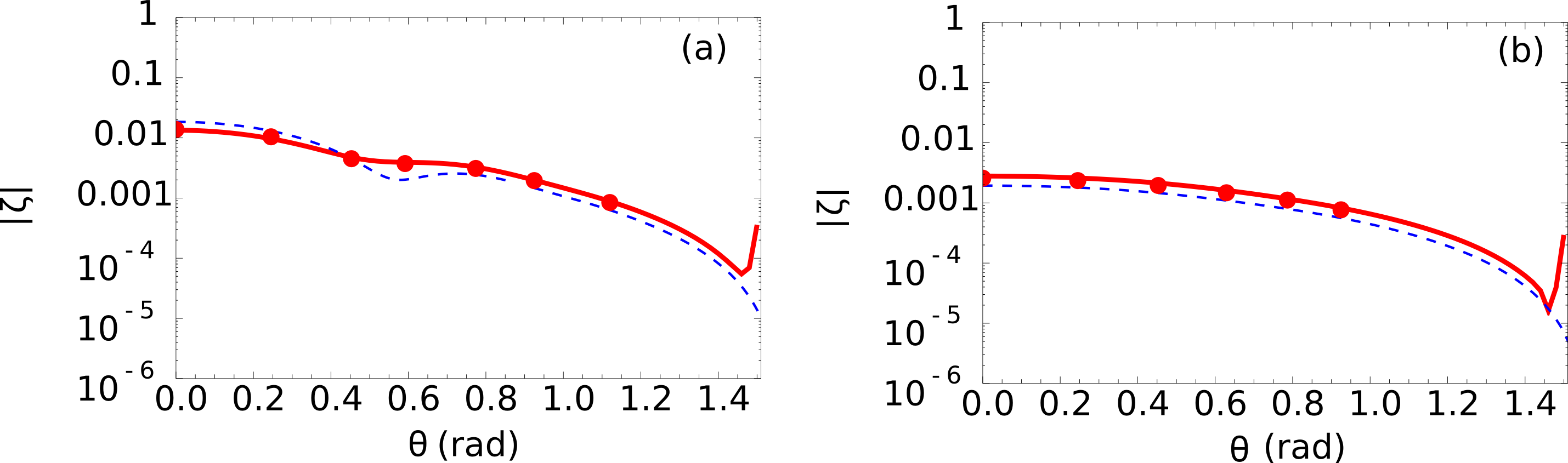}
\caption{Variation of the total re-emission coefficient of a PML $|\zeta|$ with the angle of incidence $\theta$. Fixed parameters are $\eta=\delta t/\delta x=0.4$, $\phi=0 $  and order $p=64$ (a) Variation of $|\zeta|$ with $\theta$ (in rad) for $\lambda/\delta x=4$. The red line corresponds to $|\zeta|$ computed with our "multi-sources" model and the red points to $|\zeta_{sim}|$ measured in simulations. The blue dashed line corresponds to $|\zeta|$ computed with the "1-source" model as developed in \cite{Lee2015}. (b) Variation of $|\zeta|$ with $\theta$ (in rad) for $\lambda/\delta x=8$. The red line corresponds to $|\zeta|$ computed with our "multi-sources" model and the red points to $|\zeta_{sim}|$ measured in simulations. The blue dashed line corresponds to $|\zeta|$ computed with the "1-source" model as developed in \cite{Lee2015}.}
\label{fig:PMLtheta}
\end{figure}
\end{center}

Fig. \ref{fig:PMLtheta} represents the variation of the re-emission coefficient $|\zeta|$ with the angle $\theta$ for two different wavelengths $\lambda/\delta x=4$ (Panel (a)) and $\lambda/\delta x=8$ (Panel (b)).  The solid red line corresponds to our "multi-sources" model and the dashed line to the "1-source" model. In both cases, our model perfectly matches simulation results. 

\section{Conclusion and prospects}
This paper presented a novel very general approach allowing the prediction of errors induced by any modifications of stencil in discrete solvers of Maxwell's equations on cartesian grids. The scope of this method is broad as it in principle applies to any linear system of discrete equations. In the case of electromagnetic simulations, our study demonstrates that this model can be efficiently used to predict the amount of truncation errors coming from the use of domain decomposition technique or PML with very high-order/pseudo-spectral solvers. 

In particular, our model shows that very high order field solvers can still be used with the cartesian domain decomposition technique and a reasonably low number of guard cells $N_{guards}$ without significant loss of accuracy. The number of guard cells $N_{guards}$ so that truncation errors $\zeta$ do not spoil the required accuracy in the simulation can now be predicted accurately with the new model.  

Besides, the general formalism provided by this approach allows the implementation of any configuration and the testing of arbitrary domain decomposition technique at arbitrary order. For instance, this model may be used before running a large 3D parallel electromagnetic simulation to predict the total amount of truncation errors coming from the use of domain decomposition/PML and choose the optimal set of parameters to obtain the final solution with a given accuracy. This is of great interest as running our model will be vastly faster than running the actual 3D simulation to effectively measure truncation errors and adapt the numerical parameters afterwards.

 \section*{Acknowledgement}
 
We are thankful to Brendan Godfrey for thoughtful discussions and his careful reading of the drafts leading to this paper. This work was supported by the European Commission through the Marie Slowdoska-Curie actions (Marie Curie IOF fellowship PICSSAR grant number 624543) as well as by the Director, Office of Science, Office of High Energy Physics, U.S. Dept. of Energy under Contract No. DE-AC02-05CH11231, and US-DOE SciDAC program ComPASS..

This document was prepared as an account of work sponsored in part
by the United States Government. While this document is believed to
contain correct information, neither the United States Government
nor any agency thereof, nor The Regents of the University of California,
nor any of their employees, nor the authors makes any warranty, express
or implied, or assumes any legal responsibility for the accuracy,
completeness, or usefulness of any information, apparatus, product,
or process disclosed, or represents that its use would not infringe
privately owned rights. Reference herein to any specific commercial
product, process, or service by its trade name, trademark, manufacturer,
or otherwise, does not necessarily constitute or imply its endorsement,
recommendation, or favoring by the United States Government or any
agency thereof, or The Regents of the University of California. The
views and opinions of authors expressed herein do not necessarily
state or reflect those of the United States Government or any agency
thereof or The Regents of the University of California.

\newpage
\bibliographystyle{ieeetr}

\newpage
\appendix 

\section{From arbitrary-order solvers to pseudo-spectral solvers}

We briefly review the arbitrary-order staggered Maxwell's scheme which is amongst the most commonly used solvers in electromagnetic simulations. 

We then verify analytically that when the order $p$ of this scheme tends to infinity, the arbitrary order solver converges to a pseudo-spectral solver \cite{PSTDLiu}.

\subsection{Arbitrary order scheme}

Discretized Maxwell's equations in space and time in vacuum can be written in the following general form: 
\begin{eqnarray}
\textbf{E}_{\textbf{r}}^{n+1}&=&\textbf{E}_{\textbf{r}}^{n}-c\delta t \nabla_p^{*}\times c\textbf{B}_{\textbf{s}}^{n+\frac{1}{2}}\label{eqE:gen}\\
c\textbf{B}_{\textbf{s}}^{n+\frac{1}{2}}&=&c\textbf{B}_{\textbf{s}}^{n-\frac{1}{2}}-c\delta t \nabla_p^{*}\times\textbf{E}_{\textbf{r}}^{n}\label{eqB:gen}
\end{eqnarray} 

where $n$ is the time step,  \textbf{E} the electric field and \textbf{B} the magnetic field, \textbf{r} is the discrete grid on which components of the electric field \textbf{E} are defined, \textbf{s} is the discrete grid on which components of the magnetic field \textbf{B} are defined and $\nabla_p^{*}$ is the discrete finite difference operator of order $p$. For centered schemes components of  \textbf{E} and \textbf{B} are defined at the same grid points but are shifted spatially for staggered schemes. 

For instance, if we take the simple 1D case of a linearly polarized $(E_y,B_z)$ wave propagating along the $x-axis$, we get the following equations for an arbitrary order staggered scheme of order $p$: 
\begin{eqnarray}
E_{y,i}^{n+1}&=&E_{y,i}^{n}-\frac{c\delta t}{\delta x} \sum_{l=1}^{\frac{p}{2}}C_{l}^{p}\left[cB_{z,i+\frac{1}{2}+(l-1)}^{n+\frac{1}{2}}-cB_{z,i+\frac{1}{2}-l}^{n+1_2}\right] \\
cB_{z,i+\frac{1}{2}}^{n+\frac{1}{2}}&=&cB_{z,i+\frac{1}{2}}^{n-\frac{1}{2}}-\frac{c\delta t}{\delta x} \sum_{l=1}^{\frac{p}{2}}C_{l}^{p}\left[E_{y,i+l}^{n}-E_{y,i-(l-1)}^{n}\right]
\end{eqnarray} 

where $i$ is the discrete position along the $x$-axis, $\delta x$ the mesh size and $C_{l}^{p}$ the coefficient of the finite difference operator $\nabla_p^{*}$ for staggered schemes. These coefficients were first heuristically derived by Fonberg \cite{Fornberg:1990}.  Using the approach developed in \cite{Khan1999179,Khan2003303}, it is possible to derive a closed-form for the $C_{l}^{p}$ coefficients in the particular case of the staggered scheme: 

\begin{equation}
C_{l}^{p}=\frac{(-1)^{l+1}16^{1-\frac{p}{2}}(p-1)!^2}{(2l-1)^2(\frac{p}{2}+l-1)!(\frac{p}{2}-l)!(\frac{p}{2}-1)!^2}
\end{equation}

When $p\rightarrow \infty$ this yields: 

\begin{equation}
C_{l}^{\infty}=-\frac{4}{\pi}\frac{(-1)^l}{(2l-1)^2}
\end{equation}

\subsection{Infinite limit: pseudo-spectral solver}

When the order $p\rightarrow\infty$, the arbitrary order solvers converges to a pseudo-spectral solver. Fourier transforming equations (\ref{eqE:gen}, \ref{eqB:gen}) with respect to spatial coordinates yields: 

\begin{eqnarray}
\hat{\textbf{E}}^{n+1}&=&\hat{\textbf{E}}^{n}-c\delta t \hat{\nabla}_p^{*}. c\hat{\textbf{B}}^{n+\frac{1}{2}}\label{eqEFourier:gen}\\
c\hat{\textbf{B}}^{n+\frac{1}{2}}&=&c\hat{\textbf{B}}^{n-\frac{1}{2}}-c\delta t \hat{\nabla}_p^{*}. \hat{\textbf{E}}^{n}\label{eqEFourier:gen}\\\label{eqBFourier:gen}
\end{eqnarray} 

with $\hat{\textbf{E}}$, $\hat{\textbf{B}}$ and $\hat{\nabla}_p^{*}$ the Fourier transform of $\textbf{E}$, $\textbf{B}$ and $\nabla_p^{*}$. Notice that the spatial staggering of electromagnetic components is implicitly included in the expression of $\hat{\nabla}_p^{*}$.  Each component $\hat{\nabla}_{p,m}^{*}|_{m=(x,y,z)}$ is simply given by: 
\begin{equation}
\hat{\nabla}_{p,m}^{*}(k_m)= \frac{1}{(1-l_s)2\delta_m}\sum_{l=1}^{\frac{p}{2}}C_{l}^{p}\left[e^{-j_c k_m (l-l_s)\delta_m}-e^{j_c k_m (l-l_s)\delta_m}\right] 
\end{equation}

where $l_s=1/2$ if the components of the electromagnetic fields in the direction of the finite difference are staggered ($l_s=0$ otherwise) and $j_c^2=-1$. When $p\rightarrow\infty$, we find: 
\begin{equation}
\hat{\nabla}_{p,m}^{*}(k_m)= \frac{1}{(1-l_s)2\delta_m}\sum_{l=1}^{\infty}C_{l}^{\infty}\left[e^{-j_c k_m (l-l_s)\delta_m}-e^{j_c k_m (l-l_s)\delta_m}\right] 
\end{equation}

In the case of the staggered scheme, we get: 
\begin{equation}
\hat{\nabla}_{\infty,m}^{*}(k_m)= -\frac{8j}{\pi \delta_m}\sum_{l=1}^{\infty}\frac{(-1)^l}{(2l-1)^2}\sin\frac{k_m\delta_m (2l-1)}{2}
\label{fourier:staggered}
\end{equation}
which is the Fourier series development of a triangle wave. For $|k_m\delta_m|<\pi$, this yields:
\begin{equation}
\hat{\nabla}_{\infty}^{*}(\textbf{k})= j_c \textbf{k}
\end{equation}

and Maxwell's equations write: 
\begin{eqnarray}
\hat{\textbf{E}}^{n+1}&=&\hat{\textbf{E}}^{n}-j_c\textbf{k}c\delta t\hat{\textbf{B}}^{n+1/2}\label{eqEFourier:gen}\\
c\hat{\textbf{B}}^{n+1/2}&=&c\hat{\textbf{B}}^{n-1/2}-j_c\textbf{k}c\delta t. \hat{\textbf{E}}^{n}\label{eqEFourier:gen}\\\label{eqBFourier:gen}
\end{eqnarray} 

which are the equations of the PSTD staggered scheme \cite{PSTDLiu}. Our study at arbitrary order $p$ is thus very general as it will also give us information on the behavior of pseudo-spectral solvers when $p\rightarrow \infty$ in presence of stencil modifications. 

\newpage

\section{Analytical approximate of the total re-emission coefficient $\zeta$}
\label{AnalyticalApproximate}
Here we provide an accurate analytical approximate of the total re-emission coefficient $\zeta$  for $\theta=0$ and $\phi=0$ (1D case). The approach can be easily generalized to the 3D case (not detailed here). 

\subsection{Finite order $p$}
\label{anapproxp}
By noticing that stencil coefficients $C_l^p$ decrease fast with $l$, we can infer that the pseudo-sources amplitudes $\xi_l$ due to stencil truncations will also decrease fast from the subdomain boundary where the maximum truncation occurs. Considering only a finite number $n_{\zeta}<2 (p-2)$ of pseudo-sources (instead of the required $2(p-2)$ pseudo-sources at order $p$ to get the exact analytical solution) in the calculation of $\zeta$ should thus yield an accurate analytical approximate of $\zeta$. Practically, we checked  that using $n_{\zeta}=3$ pseudo-sources $\xi_{-\frac{1}{2}}$, $\xi_{0}$ and $\xi_{\frac{1}{2}}$  in the calculation of $\zeta$ (cf. equation (\ref{eqr})) yields accurate estimate of $\zeta$:

\begin{equation}
\zeta=-\xi_{-\frac{1}{2}}e^{-j_ck\delta x}+\xi_{0}-\xi_{\frac{1}{2}}e^{-j_ck\delta x}
\label{analyticalapprox}
\end{equation}

where $\xi_{-\frac{1}{2}}$, $\xi_{0}$ and $\xi_{\frac{1}{2}}$ are solution of equation (\ref{eqmod}) with $A_{m}=(a_{i,j})_{1\leqslant i\leqslant3,1\leqslant j\leqslant3}$ and $B=(b_{i})_{1\leqslant i \leqslant3}$ given by: 

\begin{eqnarray}
a_{1,1}&=&-2\eta_x\sum_{l=1}^{\frac{p}{2}}C_l^p\cos(l-1)k\delta x+\eta_x\sum_{l=N_{guards}+1}^{\frac{p}{2}}C_l^p e^{-j_c(l-1)k\delta x}\\
a_{1,2}&=&2j_c\eta_x\sum_{l=2}^{\frac{p}{2}}C_l^p\sin(l-1)k\delta x+\eta_x\sum_{l=N_{guards}+1}^{p/2}C_l^p e^{-j_c(l-1)k\delta x}\\
a_{1,3}&=&-2\eta_x\sum_{l=2}^{\frac{p}{2}}C_l^p\cos(l-1)k\delta x+\eta_x\sum_{l=N_{guards}+1}^{p/2}C_l^p e^{-j_c(l-1)k\delta x}\\
a_{2,1}&=&1-e^{j_c\omega\delta t}\\
a_{2,2}&=&1-e^{j_c\omega\delta t}-2e^{j_c\omega\delta t/2}\eta_x\sum_{l=1}^{p/2}C_l^p e^{-j_c(2l-1)k\delta x/2}+\eta_x e^{j_c\omega\delta t/2}\sum_{l=N_{guards}+1}^{p/2}C_l^p e^{-j_c(2l-1)k\delta x/2}\\
a_{2,3}&=&2j_ce^{j_c\omega\delta t/2}\left(\sin\frac{\omega\delta t}{2}-\eta_x\sin\frac{k \delta x}{2}C_1^p\right)\\
a_{3,1}&=&-2\eta_x\sum_{l=2}^{\frac{p}{2}}C_l^p\cos(l-1)k\delta x+\eta_x\sum_{l=N_{guards}+1}^{\frac{p}{2}}C_l^p e^{-j_c(l-1)k\delta x}\\
a_{3,2}&=&-2j_c\eta_x\sum_{l=2}^{\frac{p}{2}}C_l^p\sin(l-1)k\delta x-\eta_x\sum_{l=N_{guards}+1}^{\frac{p}{2}}C_l^p e^{-j_c(l-1)k\delta x}\\
a_{3,3}&=&-2\eta_x\sum_{l=1}^{\frac{p}{2}}C_l^p\cos l k\delta x-\eta_x\sum_{l=N_{guards}+1}^{\frac{p}{2}}C_l^p e^{-j_c(l-1)k\delta x}
\end{eqnarray}

and: 

\begin{eqnarray}
b_{1}&=&\eta_x\sum_{l=N_{guards}+1}^{\frac{p}{2}}C_l^p e^{-j_c(l-1)k\delta x}\\
b_{2}&=&-\eta_x j_c e^{j_c\omega\delta t/2}\sum_{l=N_{guards}+1}^{p/2}C_l^p \sin(2l-1)\frac{k\delta x}{2}\\
b_{3}&=&-\eta_x\sum_{l=N_{guards}+1}^{\frac{p}{2}}C_l^p e^{j_c(l-1)k\delta x}\\
\end{eqnarray}

where "Strategy-2" was used as domain decomposition strategy. 

Solving equation (\ref{eqmod}) yields: 

\begin{eqnarray}
\xi_{-\frac{1}{2}}&=&\frac{b_3 a_{1,3} a_{2,2}-b_1 a_{3,3} a_{2,2}-b_3 a_{1,2} a_{2,3}-b_2 a_{1,3} a_{3,2}+b_1 a_{2,3} a_{3,2}+b_2 a_{1,2} a_{3,3}}{a_{1,3} a_{2,2} a_{3,1}-a_{1,2}
   a_{2,3} a_{3,1}-a_{1,3} a_{2,1} a_{3,2}+a_{1,1} a_{2,3} a_{3,2}+a_{1,2} a_{2,1} a_{3,3}-a_{1,1} a_{2,2} a_{3,3}}\label{eqzetam1s2}\\
\xi_{0}&=&\frac{b_3 a_{1,3} a_{2,1}-b_1 a_{3,3} a_{2,1}-b_3 a_{1,1} a_{2,3}-b_2 a_{1,3} a_{3,1}+b_1 a_{2,3} a_{3,1}+b_2 a_{1,1} a_{3,3}}{-a_{1,3} a_{2,2} a_{3,1}+a_{1,2}
   a_{2,3} a_{3,1}+a_{1,3} a_{2,1} a_{3,2}-a_{1,1} a_{2,3} a_{3,2}-a_{1,2} a_{2,1} a_{3,3}+a_{1,1} a_{2,2} a_{3,3}}\label{eqzeta0}\\
\xi_{\frac{1}{2}}&=&\frac{b_3 a_{1,2} a_{2,1}-b_1 a_{3,2} a_{2,1}-b_3 a_{1,1} a_{2,2}-b_2 a_{1,2} a_{3,1}+b_1 a_{2,2} a_{3,1}+b_2 a_{1,1} a_{3,2}}{a_{1,3} a_{2,2} a_{3,1}-a_{1,2}
   a_{2,3} a_{3,1}-a_{1,3} a_{2,1} a_{3,2}+a_{1,1} a_{2,3} a_{3,2}+a_{1,2} a_{2,1} a_{3,3}-a_{1,1} a_{2,2} a_{3,3}}\label{eqzeta1s2}
\end{eqnarray}

Replacing expressions of $\xi_{-\frac{1}{2}}$, $\xi_{0}$ and $\xi_{\frac{1}{2}}$ given by equations (\ref{eqzetam1s2}), (\ref{eqzeta0}) and (\ref{eqzeta1s2}) in equation (\ref{analyticalapprox}) gives an accurate analytical approximate of $\zeta$. 

\subsection{Infinite order $p\rightarrow \infty$}
\label{anapproxinfinite}
When $p\rightarrow \infty$, we have: 

\begin{equation}
C_{l}^{p}\rightarrow C_{l}^{\infty}=-\frac{4}{\pi}\frac{(-1)^l}{(2l-1)^2}
\end{equation}

and the elements of matrix $A_{m}$ and vector $B$ can be written as: 

\begin{eqnarray*}
a_{1,1}^{\infty}&=&\frac{\eta_x  (-1)^{\text{N}_{guards}} e^{-2 i \pi  \text{$\delta $x} \text{N}_{guards}} \Phi_{L}
   \left(-e^{-2 i \pi  \text{$\delta $x}},2,\text{N}_{guards}+\frac{1}{2}\right)}{\pi
   }+\frac{\eta_x  \left(-\Phi_{L} \left(-e^{-2 i \pi  \text{$\delta
   $x}},2,\frac{1}{2}\right)-\Phi_{L} \left(-e^{2 i \pi  \text{$\delta
   $x}},2,\frac{1}{2}\right)\right)}{\pi }\\
a_{1,2}^{\infty}&=&\frac{\eta_x  (-1)^{\text{N}_{guards}} e^{-2 i \pi  \text{$\delta $x} \text{Ng}} \Phi_{L}
   \left(-e^{-2 i \pi  \text{$\delta $x}},2,\text{N}_{guards}+\frac{1}{2}\right)}{\pi
   }...\\
   ...&-&\frac{e^{-2 i \pi  \text{$\delta $x}} \eta_x  \left(e^{4 i \pi  \text{$\delta
   $x}} \Phi_{L} \left(-e^{2 i \pi  \text{$\delta $x}},2,\frac{3}{2}\right)-\Phi_{L}
   \left(-e^{-2 i \pi  \text{$\delta $x}},2,\frac{3}{2}\right)\right)}{\pi }\\
a_{1,3}^{\infty}&=&\frac{\eta_x  (-1)^{\text{N}_{guards}} e^{-2 i \pi  \text{$\delta $x} \text{N}_{guards}} \Phi_{L}
   \left(-e^{-2 i \pi  \text{$\delta $x}},2,\text{N}_{guards}+\frac{1}{2}\right)}{\pi
   }...\\
   ...&+&\frac{e^{-2 i \pi  \text{$\delta $x}} \eta_x  \left(\Phi_{L} \left(-e^{-2 i \pi 
   \text{$\delta $x}},2,\frac{3}{2}\right)+e^{4 i \pi  \text{$\delta $x}} \Phi_{L}
   \left(-e^{2 i \pi  \text{$\delta $x}},2,\frac{3}{2}\right)\right)}{\pi }\\
a_{2,1}^{\infty}&=&1-e^{j_c\omega\delta t}\\
a_{2,2}^{\infty}&=&-\frac{8 \eta_x  e^{\frac{i \text{$\delta $t} \omega }{2}-i \pi  \text{$\delta
   $x}} \, _3F_2\left(\frac{1}{2},\frac{1}{2},1;\frac{3}{2},\frac{3}{2};-e^{-2 i
   \pi  \text{$\delta $x}}\right)}{\pi }-e^{i \text{$\delta $t} \omega
   }...\\
   ...&+&\frac{\eta_x  (-1)^{\text{N}_{guards}} \Phi_{L} \left(-e^{-2 i \pi  \text{$\delta
   $x}},2,\text{N}_{guards}+\frac{1}{2}\right) e^{\frac{i \text{$\delta $t} \omega
   }{2}-i \pi  \text{$\delta $x} (2 \text{N}_{guards}+1)}}{\pi }+1\\
a_{2,3}^{\infty}&=&2j_ce^{j_c\omega\delta t/2}\left(\sin\frac{\omega\delta t}{2}-\eta_x\sin\frac{k \delta x}{2}C_1^{\infty}\right)\\
a_{3,1}^{\infty}&=&\frac{\eta_x  (-1)^{\text{N}_{guards}} e^{-2 i \pi  \text{$\delta $x} \text{N}_{guards}} \Phi_{L}
   \left(-e^{-2 i \pi  \text{$\delta $x}},2,\text{N}_{guards}+\frac{1}{2}\right)}{\pi
   }...\\
   ...&+&\frac{e^{-2 i \pi  \text{$\delta $x}} \eta  \left(\Phi_{L} \left(-e^{-2 i \pi 
   \text{$\delta $x}},2,\frac{3}{2}\right)+e^{4 i \pi  \text{$\delta $x}} \Phi_{L}
   \left(-e^{2 i \pi  \text{$\delta $x}},2,\frac{3}{2}\right)\right)}{\pi }\\
a_{3,2}^{\infty}&=&\frac{e^{-2 i \pi  \text{$\delta $x}} \eta_x  \left(e^{4 i \pi  \text{$\delta $x}}
   \Phi_{L} \left(-e^{2 i \pi  \text{$\delta $x}},2,\frac{3}{2}\right)-\Phi_{L}
   \left(-e^{-2 i \pi  \text{$\delta $x}},2,\frac{3}{2}\right)\right)}{\pi
   }...\\
   ...&-&\frac{\eta_x  (-1)^{\text{N}_{guards}} e^{-2 i \pi  \text{$\delta $x} \text{N}_{guards}} \Phi_{L}
   \left(-e^{-2 i \pi  \text{$\delta $x}},2,\text{N}_{guards}+\frac{1}{2}\right)}{\pi }\\
a_{3,3}^{\infty}&=&\frac{\eta_x  (-1)^{\text{N}_{guards}} e^{-2 i \pi  \text{$\delta $x} \text{N}_{guards}} \Phi_{L}
   \left(-e^{-2 i \pi  \text{$\delta $x}},2,\text{N}_{guards}+\frac{1}{2}\right)}{\pi
   }+\frac{\eta_x  \left(-\Phi_{L} \left(-e^{-2 i \pi  \text{$\delta
   $x}},2,\frac{1}{2}\right)-\Phi_{L} \left(-e^{2 i \pi  \text{$\delta
   $x}},2,\frac{1}{2}\right)\right)}{\pi }
\end{eqnarray*}

and: 

\begin{eqnarray*}
b_{1}^{\infty}&=&\frac{\eta_x  (-1)^{\text{N}_{guards}} e^{-2 i \pi  \text{$\delta $x} \text{N}_{guards}} \Phi_{L}
   \left(-e^{-2 i \pi  \text{$\delta $x}},2,\text{N}_{guards}+\frac{1}{2}\right)}{\pi }\\
b_{2}^{\infty}&=&-\frac{\eta_x  (-1)^{\text{N}_{guards}} \left(e^{4 i \pi  \text{$\delta $x} \text{N}_{guards}}
   \Phi_{L} \left(-e^{2 i \pi  \text{$\delta
   $x}},2,\text{N}_{guards}+\frac{1}{2}\right)-\Phi_{L} \left(-e^{-2 i \pi  \text{$\delta
   $x}},2,\text{N}_{guards}+\frac{1}{2}\right)\right) e^{\frac{i \text{$\delta $t}
   \omega }{2}-2 i \pi  \text{$\delta $x} \text{N}_{guards}}}{2 \pi }\\
b_{3}^{\infty}&=&-\frac{\eta_x  (-1)^{\text{N}_{guards}} e^{2 i \pi  \text{$\delta $x} \text{N}_{guards}} \Phi_{L}
   \left(-e^{2 i \pi  \text{$\delta $x}},2,\text{N}_{guards}+\frac{1}{2}\right)}{\pi }\\
\end{eqnarray*}

where: 
\begin{enumerate}[(i)]
\item $\Phi_{L}(z,s,a)$ is the Lerch transcendent defined as  $\Phi_{L}(z,s,a)=\sum_{k=0}^{\infty}z^k(k+a)^{-s}$. This function can be evaluated to arbitrary numerical precision in Mathematica,
\item $_{p}F_q(a,b,z)$ is the generalized hypergeometric function which  can be evaluated to arbitrary numerical precision in Mathematica. 
\end{enumerate}

Solving equation (\ref{eqmod}) at infinite order then gives: 

\begin{eqnarray}
\xi_{-\frac{1}{2}}^{\infty}&=&\frac{b_3^{\infty} a_{1,3}^{\infty} a_{2,2}^{\infty}-b_1^{\infty} a_{3,3}^{\infty} a_{2,2}^{\infty}-b_3^{\infty} a_{1,2}^{\infty} a_{2,3}^{\infty}-b_2^{\infty} a_{1,3}^{\infty} a_{3,2}^{\infty}+b_1^{\infty} a_{2,3}^{\infty} a_{3,2}^{\infty}+b_2^{\infty} a_{1,2}^{\infty} a_{3,3}^{\infty}}{a_{1,3}^{\infty} a_{2,2}^{\infty} a_{3,1}^{\infty}-a_{1,2}^{\infty}
   a_{2,3}^{\infty} a_{3,1}^{\infty}-a_{1,3}^{\infty} a_{2,1}^{\infty} a_{3,2}^{\infty}+a_{1,1}^{\infty} a_{2,3}^{\infty} a_{3,2}^{\infty}+a_{1,2}^{\infty} a_{2,1}^{\infty} a_{3,3}^{\infty}-a_{1,1}^{\infty} a_{2,2}^{\infty} a_{3,3}^{\infty}}\label{eqzetam1s2}\\
\xi_{0}^{\infty}&=&\frac{b_3^{\infty} a_{1,3}^{\infty} a_{2,1}^{\infty}-b_1^{\infty} a_{3,3}^{\infty} a_{2,1}^{\infty}-b_3^{\infty} a_{1,1}^{\infty} a_{2,3}^{\infty}-b_2^{\infty} a_{1,3}^{\infty} a_{3,1}^{\infty}+b_1^{\infty} a_{2,3}^{\infty} a_{3,1}^{\infty}+b_2^{\infty} a_{1,1}^{\infty} a_{3,3}^{\infty}}{-a_{1,3}^{\infty} a_{2,2}^{\infty} a_{3,1}^{\infty}+a_{1,2}^{\infty}
   a_{2,3}^{\infty} a_{3,1}^{\infty}+a_{1,3}^{\infty} a_{2,1}^{\infty} a_{3,2}^{\infty}-a_{1,1}^{\infty} a_{2,3}^{\infty} a_{3,2}^{\infty}-a_{1,2}^{\infty} a_{2,1}^{\infty} a_{3,3}^{\infty}+a_{1,1}^{\infty} a_{2,2}^{\infty} a_{3,3}^{\infty}}\label{eqzeta0}\\
\xi_{\frac{1}{2}}^{\infty}&=&\frac{b_3^{\infty} a_{1,2}^{\infty} a_{2,1}^{\infty}-b_1^{\infty} a_{3,2}^{\infty} a_{2,1}^{\infty}-b_3^{\infty} a_{1,1}^{\infty} a_{2,2}^{\infty}-b_2^{\infty} a_{1,2}^{\infty} a_{3,1}^{\infty}+b_1^{\infty} a_{2,2}^{\infty} a_{3,1}^{\infty}+b_2^{\infty} a_{1,1}^{\infty} a_{3,2}^{\infty}}{a_{1,3}^{\infty} a_{2,2}^{\infty} a_{3,1}^{\infty}-a_{1,2}^{\infty}
   a_{2,3}^{\infty} a_{3,1}^{\infty}-a_{1,3}^{\infty} a_{2,1}^{\infty} a_{3,2}^{\infty}+a_{1,1}^{\infty} a_{2,3}^{\infty} a_{3,2}^{\infty}+a_{1,2}^{\infty} a_{2,1}^{\infty} a_{3,3}^{\infty}-a_{1,1}^{\infty} a_{2,2}^{\infty} a_{3,3}^{\infty}}\label{eqzeta1s2}
\end{eqnarray}

and $\zeta_{\infty}$ can then be calculated using: 

\begin{equation}
\zeta_{\infty}=-\xi_{-\frac{1}{2}}^{\infty}e^{-j_ck\delta x}+\xi_{0}^{\infty}-\xi_{\frac{1}{2}}^{\infty}e^{-j_ck\delta x}
\end{equation}

\newpage

\section{PML modeling with our multi-sources model.}

The standard formulation of the PML \cite{BerengerPML} for a p-polarized wave (with $\theta\neq0$ and $\phi=0$) requires the splitting of $B_z$ on two components $B_{zx}$ and $B_{zy}$ that verify: 

\begin{eqnarray}
B_{zx,i+\frac{1}{2},j+\frac{1}{2},k}^{n+\frac{1}{2}}&=&\alpha(i+\frac{1}{2})B_{zx,i+\frac{1}{2},j+\frac{1}{2},k}^{n-\frac{1}{2}}+\beta(i+\frac{1}{2})\sum_{l=1}^{\frac{p}{2}}C_{l}^{p}\left[E_{y,i+l,j+\frac{1}{2},k}^{n}-E_{y,i-(l-1),j+\frac{1}{2},k}^{n}\right]\\
B_{zy,i+\frac{1}{2},j+\frac{1}{2},k}^{n+\frac{1}{2}}&=&B_{zy,i+\frac{1}{2},j+\frac{1}{2},k}^{n-\frac{1}{2}}-\eta_y\sum_{l=1}^{\frac{p}{2}}C_{l}^{p}\left[ E_{x,i+\frac{1}{2},j+l,k}^{n}-E_{x,i+\frac{1}{2},j-(l-1),k}^{n}\right]
\end{eqnarray}

with $B_z=B_{zx}+B_{zy}$ and $\alpha$, $\beta$ coefficients of the PML. In the case of a plane monochromatic driving wave of frequency $\omega$, replacing the expression of $B_{zx}$ and $B_{zy}$ in $B_z$ we get the following equation for $B_z$:

\begin{flalign}
&\begin{array}{lcl}
 cB_{z,i+\frac{1}{2},j+\frac{1}{2},k}^{n+\frac{1}{2}}&=&cB_{z,i+\frac{1}{2},j+\frac{1}{2},k}^{n-\frac{1}{2}}\\
 &+&\beta(i+\frac{1}{2})\gamma(i+\frac{1}{2})\sum_{l=1}^{\frac{p}{2}}C_{l}^{p} \left[E_{y,i+l,j+\frac{1}{2},k}^{n}-E_{y,i-(l-1),j+\frac{1}{2},k}^{n}\right]\\
 &+&\eta_y \sum_{l=1}^{\frac{p}{2}}C_{l}^{p}\left[ E_{x,i+\frac{1}{2},j+l,k}^{n}-E_{x,i+\frac{1}{2},j-(l-1),k}^{n}\right] \end{array}&\label{eqBzijk2}
\end{flalign}

with: 

\begin{equation}
\gamma(i)=\frac{e^{j_c\omega\delta t/2}-e^{-j_c\omega\delta t/2}}{e^{j_c\omega\delta t/2}-\alpha(i)e^{-j_c\omega\delta t/2}}
\end{equation}

which is equivalent to take the following coefficients in the general formulation of equation (\ref{eqBzijk}):

\begin{equation}
\psi_{i,l}^{p}=\Gamma_{i,l}^{p}=C_{l,p}, \forall i
\end{equation}

and:

\begin{eqnarray}
\alpha^*_{z,i+\frac{1}{2}}&=&1\\
\beta^*_{z,i+\frac{1}{2}}&=&\beta(i+\frac{1}{2})\gamma(i+\frac{1}{2})\\
\gamma^{*}_{z,i+\frac{1}{2}}&=&\eta_{y}\\
\end{eqnarray}

\end{document}